\pdfoutput=1 
\documentclass[aps,pre,preprint, superscriptaddress]{revtex4}

\usepackage{amsmath}
\usepackage{amsfonts}
\usepackage{amssymb}
\usepackage{pdfpages}
\usepackage{epstopdf}
\usepackage{graphicx}  
\usepackage{mathrsfs}
\usepackage[percent]{overpic} 
\usepackage{float}  
\usepackage{appendix} 

\DeclareGraphicsExtensions{.pdf,.eps}

\begin{document}

\title{Impact of rainfall on {\it Aedes aegypti} populations}

\author{L. D. Valdez} \affiliation{Facultad
  de Matem\'atica, Astronom\'ia, F\'isica y Computaci\'on, Universidad
  Nacional de C\'ordoba, Instituto de F\'isica Enrique Gaviola,
  CONICET, Ciudad Universitaria, 5000 C\'ordoba, Argentina}
\author{G. J. Sibona} \affiliation{Facultad
  de Matem\'atica, Astronom\'ia, F\'isica y Computaci\'on, Universidad
  Nacional de C\'ordoba, Instituto de F\'isica Enrique Gaviola,
  CONICET, Ciudad Universitaria, 5000 C\'ordoba, Argentina}
\author{C. A. Condat} \affiliation{Facultad
  de Matem\'atica, Astronom\'ia, F\'isica y Computaci\'on, Universidad
  Nacional de C\'ordoba, Instituto de F\'isica Enrique Gaviola,
  CONICET, Ciudad Universitaria, 5000 C\'ordoba, Argentina}

\begin{abstract}
{\it Aedes aegypti} is the main vector of multiple diseases, such as
Dengue, Zika, and Chikungunya. Due to modifications in weather
patterns, its geographical range is continuously evolving. Temperature
is a key factor for its expansion into regions with cool winters, but
rainfall can also have a strong impact on the colonization of these
regions, since larvae emerging after a rainfall are likely to die at
temperatures below $10^{\circ}$C. As climate change is expected to
affect rainfall regimes, with a higher frequency of heavy storms and
an increase in drought-affected areas, it is important to understand
how different rainfall scenarios may shape {\it Ae. aegypti}'s
range. We develop a model for the population dynamics of {\it
  Ae. aegypti}, coupled with a rainfall model to study the effect of
the temporal distribution of rainfall on mosquito abundance. Using a
fracturing process, we then investigate the effect of a higher
variability in the daily rainfall. As an example, we show that
rainfall distribution is necessary to explain the geographic range of
{\it Ae. aegypti} in Taiwan, an island characterized by rainy winters
in the north and dry winters in the south. We also predict that a
higher variability in the rainfall time distribution will decrease the
maximum abundance of {\it Ae. aegypti} during the summer. An increase
in daily rainfall variability will likewise enhance its extinction
probability. Finally, we obtain a nonlinear relationship between dry
season duration and extinction probability. These findings can have a
significant impact on our ability to predict disease outbreaks.
\end{abstract}
\maketitle

\section{Introduction}

The prevalence areas of both emerging and
well-established~\cite{hay2002climate,rochlin2013climate,morin2013regional,caminade2017global,xu2016climate}
vector-borne diseases are likely to be substantially modified by
climate variations. Being {\it Ae. aegypti} the main vector of several
important diseases, its ecology is currently the focus of intense
research. Since climate is a key determinant of the mosquito
habitat~\cite{hopp2001global}, climate change is expected to
significantly alter its geographic range and put new regions at
risk. To predict the future evolution of this range it is
necessary to have a clear understanding of how different climatic
factors affect {\it Ae. aegypti}'s thriving and survival.  While
temperature governs its reproduction, maturation and mortality
rates~\cite{bar1958effect}, rainfalls generate breeding grounds for
larvae and pupae~\cite{moore1978aedes}. At variance with other
mosquito species, {\it Ae. aegypti}'s eggs are laid above the water
surface and hatch only when the water level rises and wets them. The
long survival times of its dry eggs endow {\it Ae. aegypti} with a
competitive advantage over other mosquito species during long periods
of drought, but a winter rain may force their hatching and the
subsequent larval death. The determination of how climate change may
affect the geographical distribution of this mosquito is thus highly
nontrivial. In this paper we address the influence of different
rainfall regimes on {\it Ae. aegypti}.

{\it Aedes aegypti} inhabits tropical and subtropical regions
worldwide, its geographic range being roughly limited by the
$10^{\circ}$C winter isotherms~\cite{christophers1960aedes}. The areas
close to these isotherms are called ``cool
margins''~\cite{eisen2013aedes,eisen2014impact}. Although the
literature describing the reproduction, maturation, and mortality
rates, and the dynamics of the mosquito population in warm climates is
vast~\cite{tun2000effects,maciel2007daily,katyal1996seasonal,
  magori2009skeeter}, fewer studies have been conducted in the ``cool
margins'', which often receive cold fronts in winter. Rozeboom
\cite{rozeboom1939overwintering} studied the survival of {\it
  Ae. aegypti} during the winter in Stillwater, Okla. (USA), where the
temperature is often below freezing. This author found that only those
{\it Ae. aegypti} eggs that were protected from rain and snow became
vigorous adults. Later on, Tsuda and Takagi~\cite{tsuda2001survival}
studied the survival of larvae in Nagasaki, Japan, finding that they
did not tolerate the low winter temperatures. While Tsuda and Takagi
did not perform a direct study with rainfall, they suggested that
winter rainfalls could cause mosquito eggs to hatch before spring, and
hence the larvae could die due to the low temperatures. Similarly,
Chang et al.~\cite{chang2007differential} found in field studies in
Taiwan that larval mortality increases rapidly due to cold
fronts. Since rainfall may trigger the hatching process, winter
rainfalls could impact negatively on {\it Ae. aegypti}'s ability to
colonize new regions, especially in the ``cool margins''. In addition,
as climate change is expected not only to increase the temperature but
also the frequency of storms and
droughts~\cite{easterling2000climate}, it is important to evaluate how
a higher {\it variability} in precipitation will affect the dynamics of the
mosquito population.

In this work we study the effect of different rainfall regimes on the
survival of {\it Ae. aegypti}, using Taiwan as an example for our
description. This island is an excellent case study. Its average summer
temperatures are usually above $20^{\circ}$C and it receives abundant
rainfall throughout its territory \cite{Exxon_05}. Taiwan has winter
isotherms above $10^{\circ}$C, although northern temperatures are
slightly lower than those in the south. Crucially, the rainfall regime
in winter is not spatially homogeneous, due to the presence of the
Central Mountain Range~\cite{chen2003rainfall}. For instance, in
January Taipei (located in the north) receives an average rainfall of
83mm, whereas in Kaohsiung (located in the south) the average rainfall
is 16mm. On the other hand, despite the small surface of Taiwan,
entomological studies indicate that the {\it Ae. aegypti} population
occurs only in the
south~\cite{yang2014discriminable,hwang1991ecology}, so we posit that
winter rainfall is a determinant for the absence of {\it Ae. aegypti} in the
north of the country.

We calibrate our model of mosquito populations to reproduce the actual
geographical distribution of mosquitoes in four representative
Taiwanese cities. Our model correctly predicts that {\it Ae. aegypti}
should not be present in Taipei, but thrive in Kaohsiung, but it also
predicts that, if we reversed the rainfall data for these two cities
we would find that this species would become extinct in Kaohsiung but
prosper in Taipei. This result provides a strong validation of our
working hypothesis. To find out how different rainfall regimes would impact
on the mosquito population dynamics, we present a model to generate
synthetic rainfall time series. This model is based on a fracturing
method~\cite{finley2014exploring}, which allows us to modify the
temporal distribution of rainfall. Using this rainfall model, we find
that the four Taiwanese cities have favorable conditions during the summer for
the reproduction of {\it Ae.  aegypti}. In addition, we obtain that,
as the variability of rainfall increases, the maximum mosquito
abundance $M_{max}$ diminishes. Then, we explore the effect of winter
on the survival of {\it Ae. aegypti}, obtaining that in all cities
except Taipei a decrease in rainfall variability reduces the
likelihood of mosquito extinction. Additionally, we analyze how
mosquitoes would withstand winter in the four cities if the rate of
reproduction changed considering the same rainfall regime, and we find
that Kaohsiung is still the most favorable city for these
mosquitoes. Finally, we study how the duration of the dry period
affects mosquito survival, obtaining a nonlinear relationship between
these two variables.

\section{Models and Results}

\subsection{The model of mosquito abundance}\label{Sec.modd}

\begin{figure}[H]
\centering
\begin{overpic}[scale=0.6]{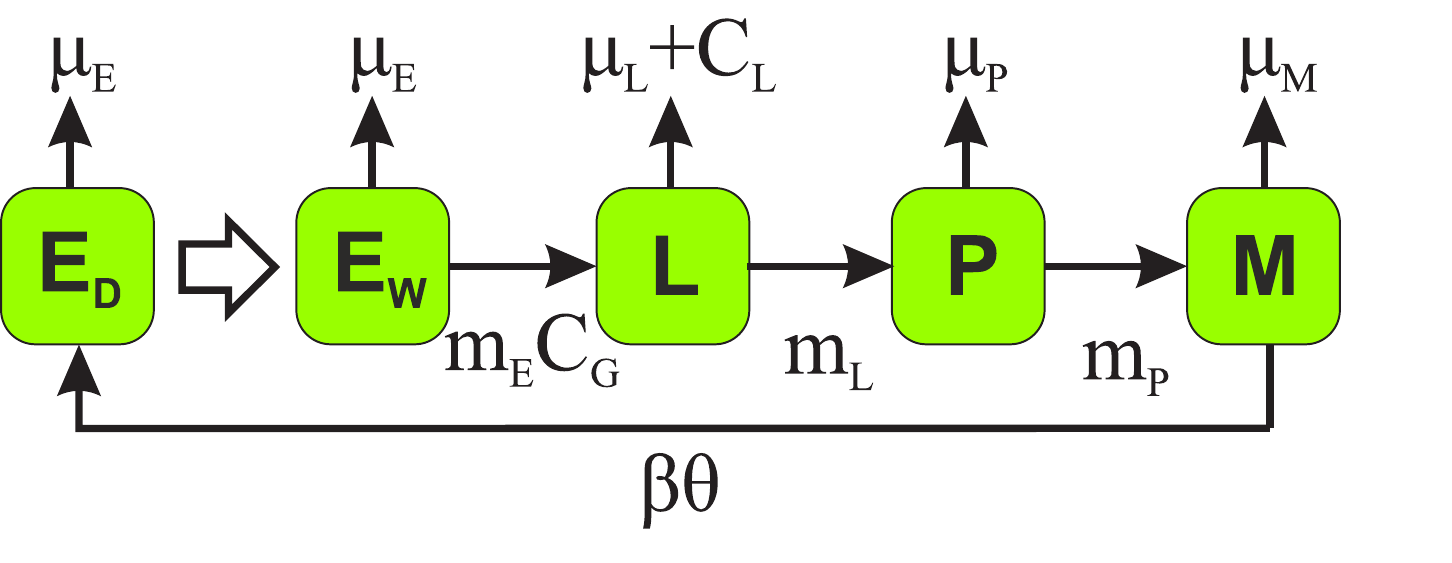}
\end{overpic}
\caption{Schematic representation of the {\it Ae. aegypti} evolution
  through different maturation stages.  Mosquito compartments are: dry
  eggs ($E_D$), wet eggs ($E_w$), larvae (L), pupae (P), and adult
  mosquitoes (M). Thin arrows symbolize a waiting time transition
  which follows an exponential function, whereas the thick arrow
  represents a delta waiting time distribution from dry eggs to wet eggs
  (see Sec.~``Rainfall and wet eggs''). Transition, maturation, and mortality
  rates are described in the Appendix.}\label{figEsq1}
\end{figure}

In this section we introduce a climate-driven abundance model of {\it
  Ae.  aegypti} mosquitoes, using four main compartments: eggs
($E_T$), larvae (L), pupae (P), and adult mosquitoes (M). In addition,
we distinguish between dry ($E_D$) and wet eggs ($E_W$). The former
are those eggs that have not been in contact with water and therefore
cannot hatch, while the latter are those that were in contact with
water, which we will assume to come only from rainfall. In our model
only wet eggs develop into larvae. In
Fig.~\ref{figEsq1} we show a schematic of the transitions between
compartments and in Table~\ref{tab.Trans} we present a summary of the
parameters of our model that are related with oviposition and
rainfall. In the following, we will explain the dependence of the
transition rate coefficients with temperature and rainfall.

\begin{table}[H]
\centering
\caption{Variables and parameters related with oviposition and rainfall}
\label{tab.Trans}
\begin{tabular}{l l}
\hline
Quantity & Definition  \\
\hline
$\beta$& birth rate of mosquitoes in optimal conditions (days$^{-1}$)  \\
$\theta$& effect of the temperature on the mosquito birth rate\\
$K_L$& carrying capacity\\
$H_{max}$&maximum daily amount of accumulated rainwater [mm]\\
$H$&accumulated amount of rainwater [mm]\\
$R$&daily rainfall [mm]\\
$Evap$&daily evapotranspiration [mm]\\
$k$& constant of the Ivanov model [mm/$^{\circ}$C$^2$] (see Eq.~\ref{eq.EvapEq})\\
$T$&average daily temperature [$^{\circ}$C]\\
$Hum$&daily relative humidity\\
\hline
\end{tabular}
\end{table}

\subsubsection{Survival of {\it Ae. aegypti} at low temperatures}\label{sec.surlow}

Contrary to other {\it Aedes} species, such as {\it Ae. albopictus} or
{\it Ae. albifasciatus}, which survive freezing
temperatures~\cite{thomas2012low,kress2016effects,garzon2013resistance},
different studies suggest that {\it Ae. aegypti} eggs only survive a
few hours at low temperatures. For example,
Hatchett~\cite{hatchett1946winter} found that only $40\%$ of {\it
  Ae. aegypti} eggs could hatch after they were immersed in water and
exposed to 1$^{\circ}$C for 24hs; Chang et
al.~\cite{chang2007differential} found that the larval mortality rate
increases rapidly when the minimum temperature is below about
10$^{\circ}$C. On the other hand, historical global collections
suggest that {\it Ae. aegypti} is distributed geographically only in
areas with winter isotherms above 10$^{\circ}$C
\cite{christophers1960aedes}. Therefore, we will assume in this work
that all mosquito compartments that are in contact with water, $i.e.$
wet eggs, larvae, and pupae, lose a 50\% of their members at the end
of any day whose minimum temperature is below 10$^{\circ}$C (see
Appendix Sec.~``Integration of the equations of our model''). Above
this threshold, the dependence of the mortality rate with the
temperature for each compartment is that given by Barmak et
al.~\cite{barmak2014modelling} (see Appendix Sec.~``Mortality
rates'').

\subsubsection{Rainfall and wet eggs}\label{sec.RainWE}
{\it Ae. aegypti} lays its eggs on the inner side of containers above
the water line~(see Ref.~\cite{goddard2016physician}). These eggs are
regarded in our model as dry eggs and when they are flooded, for
example by rainwater, they usually hatch. Therefore, following the
work of Aznar et al.~\cite{aznar2013modeling}, we propose that, if on
day $d$ it rains $R$ millimeters, the fraction of dry eggs that
become wet eggs is given by the following Hill function,
\begin{eqnarray}\label{eq.hill}
f(R)= 0.8\frac{(R/R_{thres})^5}{1+(R/R_{thres})^5}.
\end{eqnarray}
Here, $R_{thres}$ is the rainfall threshold and the prefactor $0.8$
stands for the maximum fraction of eggs that may hatch when they are
flooded. In this work, we set $R_{thres}=10$[mm]. This function is
applied at the end of day $t=d $ on the dry egg compartment, where
$t$ is a continuous variable and $d$ is a non-negative integer
(see Appendix Sec.~``Integration of the equations of our model'').

\subsubsection{Effect of rainfall on the carrying capacity}
The persistence of breeding sites is a key factor for the survival of
mosquitoes in the aquatic stage. Rainfall creates breeding sites where
larvae develop prior to becoming adult mosquitoes, and evaporation
tends to shrink these sites. Therefore, we propose that the carrying
capacity depends on the amount of available water $H(t)$, whose
variation is defined as follows:
\medskip
\begin{eqnarray}\label{eq.Ht}
H(t+1)=\left\{%
\begin{array}{ll}
0 &\;\;\;\; \text{if}\;\;\;\; H(t)+ \Delta(t) \leqslant 0\\
H_{max}  & \;\;\;\; \text{if}\;\;\;\;H(t)+\Delta(t) \geqslant H_{max}\\
H(t)+\Delta(t) &\;\;\;\; \text{otherwise}, \\
\end{array}%
\right.
\end{eqnarray}

\medskip
\noindent where $\Delta(t)=R(t)-Evap(t)$, $R(t)$ is the amount of rain on day $t$ and $Evap(t)$
is the daily evaporation. Note that $H$ can increase only up to a
maximum value $H_{max}$ since we consider that at a higher water level
the containers or breeding sites overflow. Besides, following the
Ivanov model~\cite{romanenko1961computation, valipour2014application},
we propose that the evaporation rate $Evap(t)$ is given by the
expression,
\begin{eqnarray}\label{eq.EvapEq}
Evap(t)=k(25+T(t))^2(100-Hum(t)).
\end{eqnarray}
where $ T $ is the average temperature and $ Hum $ the
humidity. Finally, we propose that the carrying capacity is given by:
\begin{equation}\label{eq.KL}
K_L(t)=K_{max}\frac{H(t)}{H_{max}}+1,
\end{equation}
where $ K_{max} $ is the maximum carrying capacity and the $1$ is
introduced to avoid divergences if $H(t) \to 0$.

For more details of our model of {\it Ae. aegypti} mosquitoes see 
Appendix Sec.~``Transitions rates''. In the following section we will
apply our mosquito model to study the case of Taiwan, using the actual
rainfall time series.

\section{Case study: Taiwan}

As mentioned in the Introduction, to study how different rainfall
regimes impact on the mosquito population it is informative to
analyze the case of Taiwan. More specifically, we will apply our model
to study mosquito populations in four cities in Taiwan: Taipei
(north), Taichung (west), Hualien (east), and Kaohsiung (south). In
Figs.~\ref{fig.DatTaiw} (see Appendix Sec. ``Weather in Taiwan'')
we show their average, minimum, and maximum daily temperatures in the
period 2010-2013, and the average monthly precipitation
$\mathscr{P}(m)$ in the period 1981-2010. Of these four cities, an
established {\it Ae. aegypti} population has been found only in
Kaohsiung. We investigate the reasons why.

In order to model the actual distribution of {\it Ae. aegypti} in
these cities, we calibrate a deterministic version of our model to the
time series of adult {\it Ae. aegypti} abundances per block in
Kaohsiung in the period January 2010-December 2012. Namely, we use a
Latin hypercube sampling (LHS) method to generate points in a
four-dimensional space defined by the vector
$(\beta,k,H_{max},K_{max})$. Additionally, since Taipei, Taichung, and
Hualien do not have permanent {\it Ae. aegypti} populations, for
each point of the LHS method we also obtain the abundance of adult
mosquitoes corresponding to each city, computing the mean square
distance between them and a null abundance mosquito series. Finally,
we select the point of the LHS that minimizes the sum of the mean
square distances for the four cities. See Appendix Sec.~``Model
calibration'' for more details of the calibration process. In
Table~\ref{tab.matAjus} we show the estimated values of the
parameters.

\begin{table}[H]
\centering
\caption{Estimated values of $(\beta,k,H_{max},K_{max})$ using a Latin
  hypercube sampling.}
\label{tab.matAjus}
\begin{tabular}{|l|c|}
\hline
parameter & values\\
\hline
$\beta$& 0.52  \\
$k$&  3.9 $10^{-5}$\\
$H_{max}$&   24\\
$K_{max}$&   212\\
\hline
\end{tabular}
\end{table}

\begin{figure}[H]
\centering
\begin{overpic}[scale=0.25]{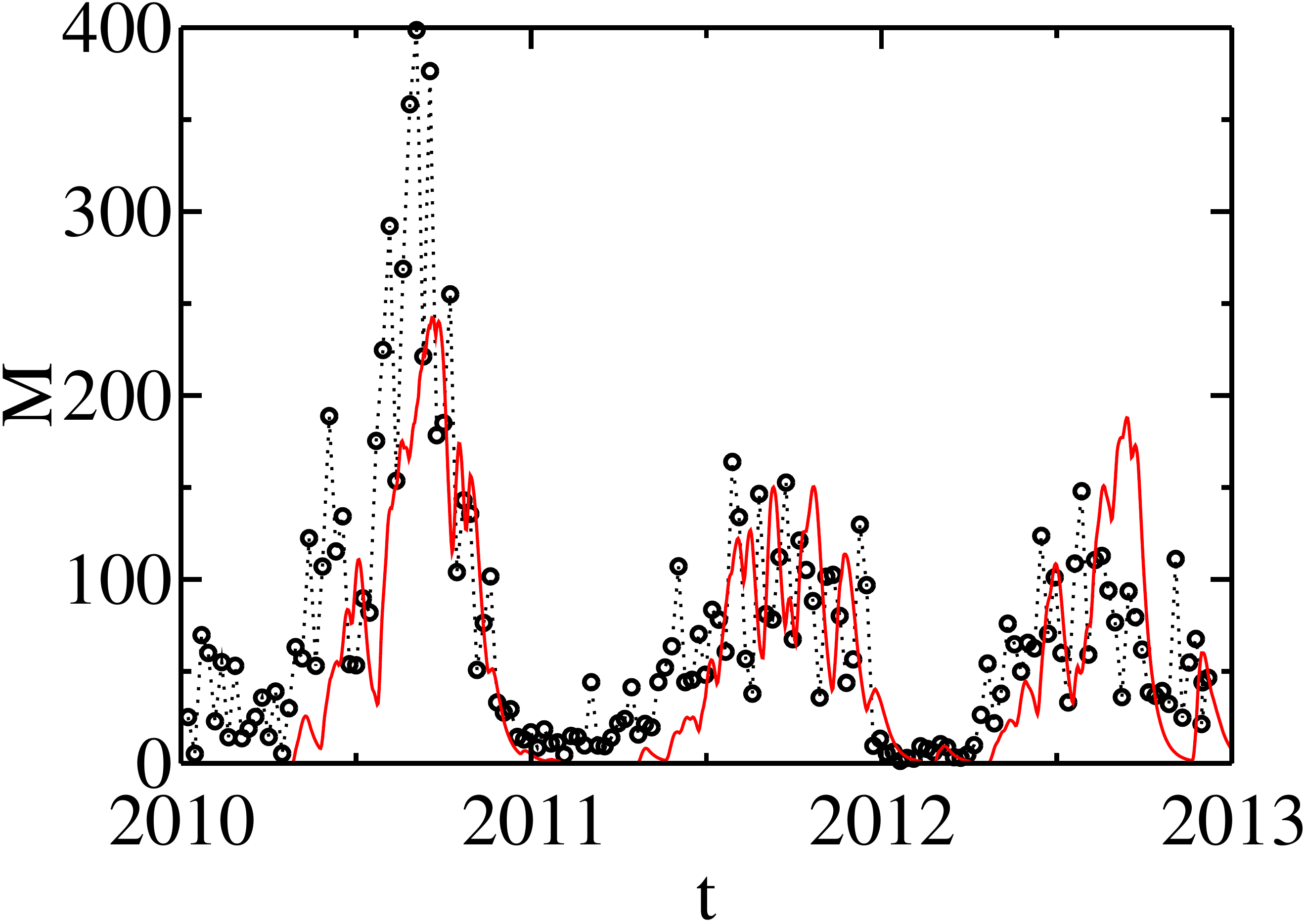}
  \put(75,50){(a)}
\end{overpic}
\begin{overpic}[scale=0.25]{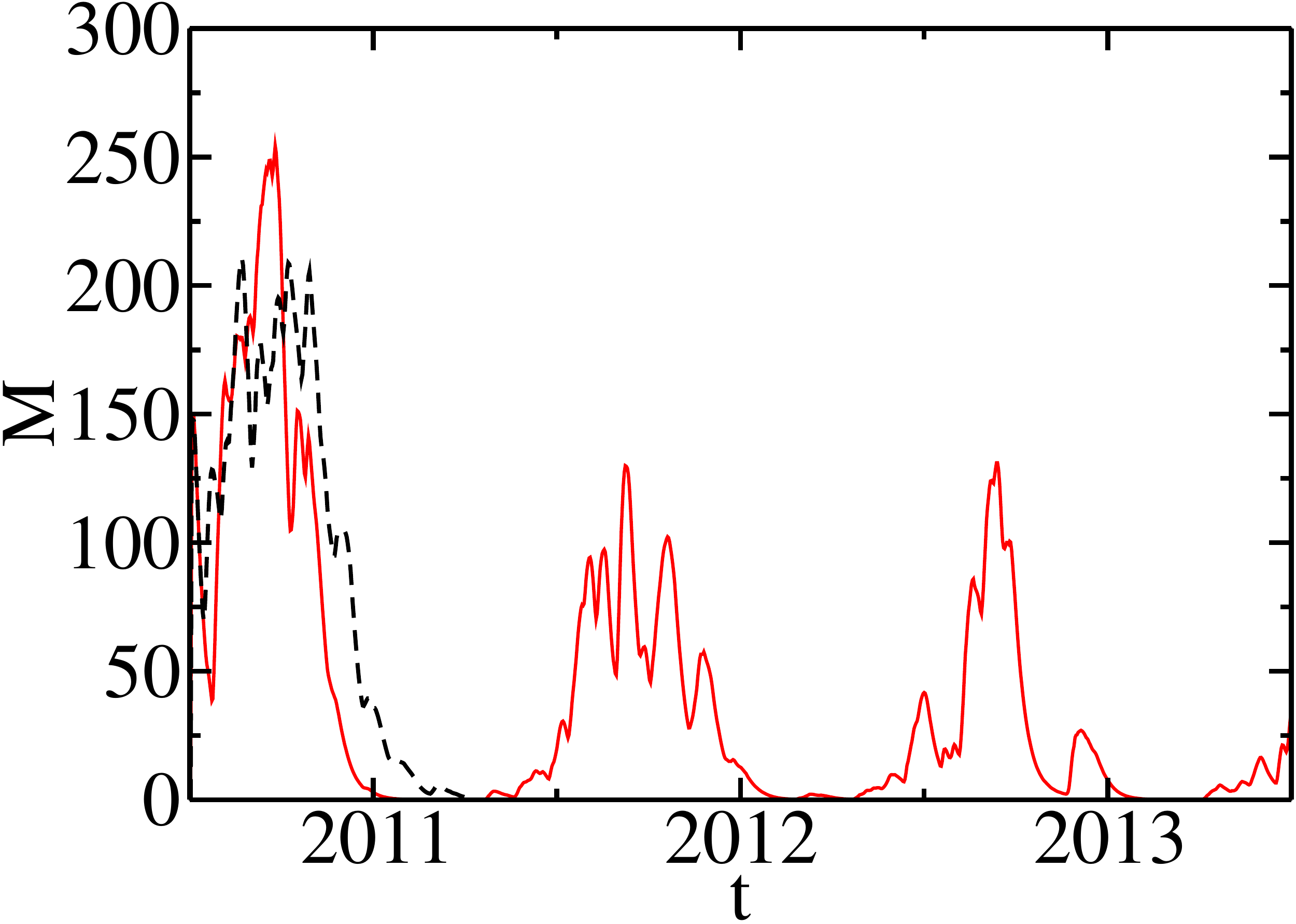}
  \put(75,50){(b)}
\end{overpic}
\caption{(a) Time series of the number of adult {\it Ae. aegypti}
  mosquitoes obtained from the Kaohsiung data
  (circles)~\cite{Exxon_04} and the prediction of our model (solid
  line), using the parameters from Table~\ref{tab.matAjus} and the
  actual time series of temperature, rainfall, and humidity for
  Kaohsiung in the period January 2010-December
  2012~\cite{Exxon_05}. For the calibration we have re-scaled the
  actual time series of adult mosquito abundances, so that this
  abundance is of 100 mosquitoes in the summer. (b) Evolution of the
  mosquito abundance using the calibrated parameters shown in
  Table~\ref{tab.matAjus} and the actual meteorological data of
  Kaohsiung and Taipei, but where we swapped the rainfall time series
  of these two cities. The solid red (dashed black) line corresponds
  to the case of temperature and humidity of Taipei (Kaohsiung) and
  the rainfall of Kaohsiung (Taipei).}\label{fig.Ajus}
\end{figure}

In Fig.~\ref{fig.Ajus}a we show the actual time series of the
abundance of adult mosquitoes~\cite{Exxon_04} and the predictions of
our model using the parameters obtained with the calibration process
and the actual time series of temperature, rainfall, and humidity of
Kaohsiung~\cite{Exxon_05}. Although our model does not reproduce
exactly the heights of the actual mosquito abundance, the positions of
the peaks are well correlated with the data. We recall that the Aedes
populations in the other cities are extinct.

In order to ascertain if rainfall is indeed responsible for the
distribution of {\it Ae. aegypti} mosquitoes in Taiwan, we computed
the dynamics of mosquito abundances in Kaohsiung and Taipei, but {\it
  exchanging the actual rainfall time series} between these places. As
it can be seen from Fig.~\ref{fig.Ajus}b, in this hypothetical
scenario where Taipei has drier winters, it could maintain a sizable
population of mosquitoes through the years. On the other hand, if
Kaohsiung had rainy winters, {\it Ae. aegypti} would become extinct
there. These results confirm that our model is effective describing
scenarios where rainfall controls the mosquito survival.

In the following sections, we introduce the model to generate
synthetic rainfall time series, which will allow us to study how
different rain regimes can shape {\it Ae. aegypti}'s geographic
distribution.

\section{Synthetic rainfall time series}\label{sec.synLL}
In order to study the dynamics of mosquito populations for different
rainfall regimes, we will use the method to generate synthetic
rainfall time series presented in Ref.~\cite{valdez2017effects}, which
we briefly review next. In this method, we must first specify the
total monthly rainfall $\mathscr{P}(m) $ and the monthly number of
rainy days $\mathscr{D}(m)$, where $m$ stands for the month. These
variables are often used in regression models to predict the number of
mosquitoes, as well as to describe the rainfall regime in
meteorological
forecasting~\cite{norrant2006monthly,modarres2007rainfall,hu2006rainfall,pascual2006malaria}. The
values of $\mathscr{P}(m)$ and $\mathscr{D}(m)$ can either be assigned
using the actual averages obtained from meteorological data, or they
can be modeled using mathematical functions.

To decide how the monthly rainfall $\mathscr{P}(m)$ for a given month
$m$ is distributed on a daily basis, we use a fracturing method which
consists in successively decomposing the monthly precipitation
$\mathscr{P}(m)$ into the sum of two terms
(see Refs.~\cite{finley2014exploring,valdez2017effects}), until
$\mathscr{P}(m)$ is the sum of $\mathscr{D}(m)$ terms. Namely, this
process begins by decomposing $\mathscr{P}(m)$ into two non-negative
terms $X_1$ and $X_2$ with $\mathscr{P}(m)=X_1+X_2$ where $X_1$ is
given by

\medskip
\begin{eqnarray}\label{alphaFTT}
X_1=\mathscr{P}(m) \times \left\{%
\begin{array}{ll}
\rho \frac{1-\alpha}{\alpha} &\;\; \text{if}\;\; 0\leqslant \rho <\frac{\alpha}{2} \\
\frac{1}{2}+\left(\rho-\frac{1}{2}\right)\frac{\alpha}{1-\alpha} &\;\; \text{if}\;\; \frac{\alpha}{2}\leqslant \rho\leqslant 1-\frac{\alpha}{2} \\
1-(1-\rho)\frac{1-\alpha}{\alpha} &\;\; \text{if}\;\; 1-\frac{\alpha}{2} < \rho \leqslant 1 \\
\end{array}%
\right.
\end{eqnarray}

\medskip
\noindent where $\rho$ is a uniform random variable and $\alpha \in [0,1] $ is a
parameter that controls the variance of the values of $X_1$. In particular, for $\alpha=0$,
$X_1=X_2=\mathscr{P}(m)/2$, while for $\alpha=1$ we obtain $X_1=0$ and
$X_2=\mathscr{P}(m)$. Therefore, an increasing in the value of
$\alpha$ generates a rainfall time series with a higher heterogeneity
in the daily amount of precipitation. This step is applied iteratively
to each term until the total number of terms is $\mathscr{D}(m)$. Each
of these terms corresponds to a total daily rainfall, which is assigned at
random to one day of the month.

Although this method allows us to explore different scenarios of
rainfall heterogeneity, it is of interest to determine the value of
$\alpha$ that best fits an actual rainfall time series, as a reference
value for our analysis. To this end, in Appendix
Sec.~``Calibration of $\alpha$'', we develop a method that computes
the value of $\alpha $ that best fits the variance of the daily
precipitation for a given period of time (usually, one month).

In Table~\ref{tab.AjAlp} we show the values of $\alpha $ for a few
cities. Notably we obtain that the values are all distributed in a
relatively narrow range ($[0.32;0.46]$), suggesting that rainfalls are
moderately heterogeneous.

\begin{table}[H]
\centering
\caption{Estimated values of $\alpha$ for several
  cities. Meteorological data were obtained from Ref.~\cite{Exxon_01} for
  the city of Hong Kong, from Ref.~\cite{Exxon_02} for the cities in
  Australia, and from Ref.~\cite{Exxon_06} for the other cities. In the second column we show the month analyzed, which
  corresponds to the rainiest month in each city.}
\label{tab.AjAlp}
\begin{tabular}{|l|c|c|}
\hline
Cities (number of years analyzed)& Month analyzed &$\bar{\alpha}$ \\
\hline
Sydney, Australia (157 years)& April &0.46\\
Hong Kong, China (126 years) & June &0.44  \\
Alice Springs, Australia (67 years)& February & 0.43 \\
Berlin, Germany (134 years) & July & 0.39\\
Atlanta, USA (88 years) & March &0.39 \\
Happy Valley, Australia (46 years)& March &0.38 \\
Los Angeles, USA (91 years) &February & 0.38 \\
London, UK (43 years) & October & 0.36\\
Chapingo, Mexico (44 years) & July &0.32\\
\hline
\end{tabular}
\end{table}

\subsection{Results with synthetic rainfall time series}\label{sec.Discussion}

\begin{figure}[H]
\centering
\begin{overpic}[scale=0.25]{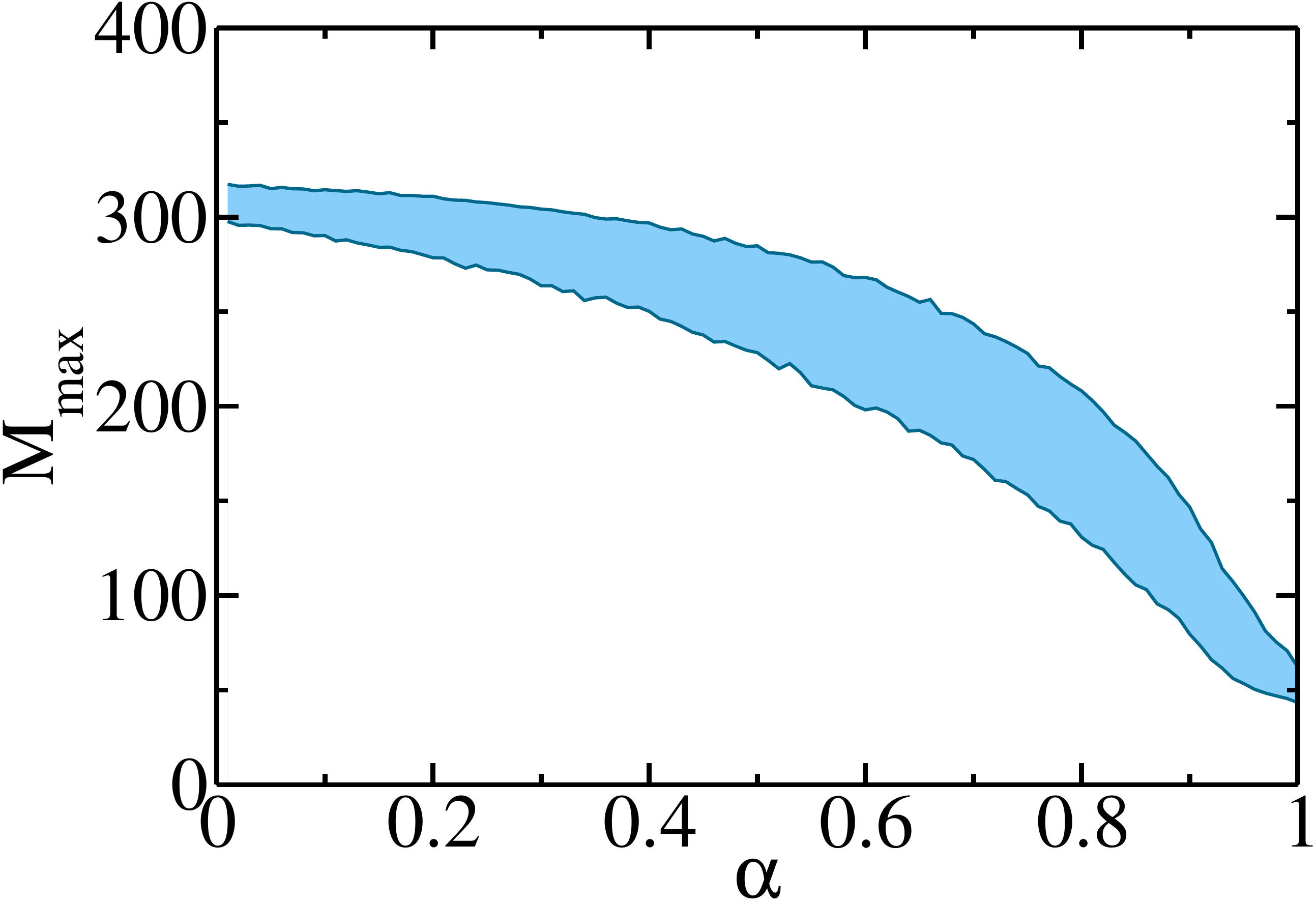}
  \put(50,55){Kaohsiung}
\end{overpic}
\begin{overpic}[scale=0.25]{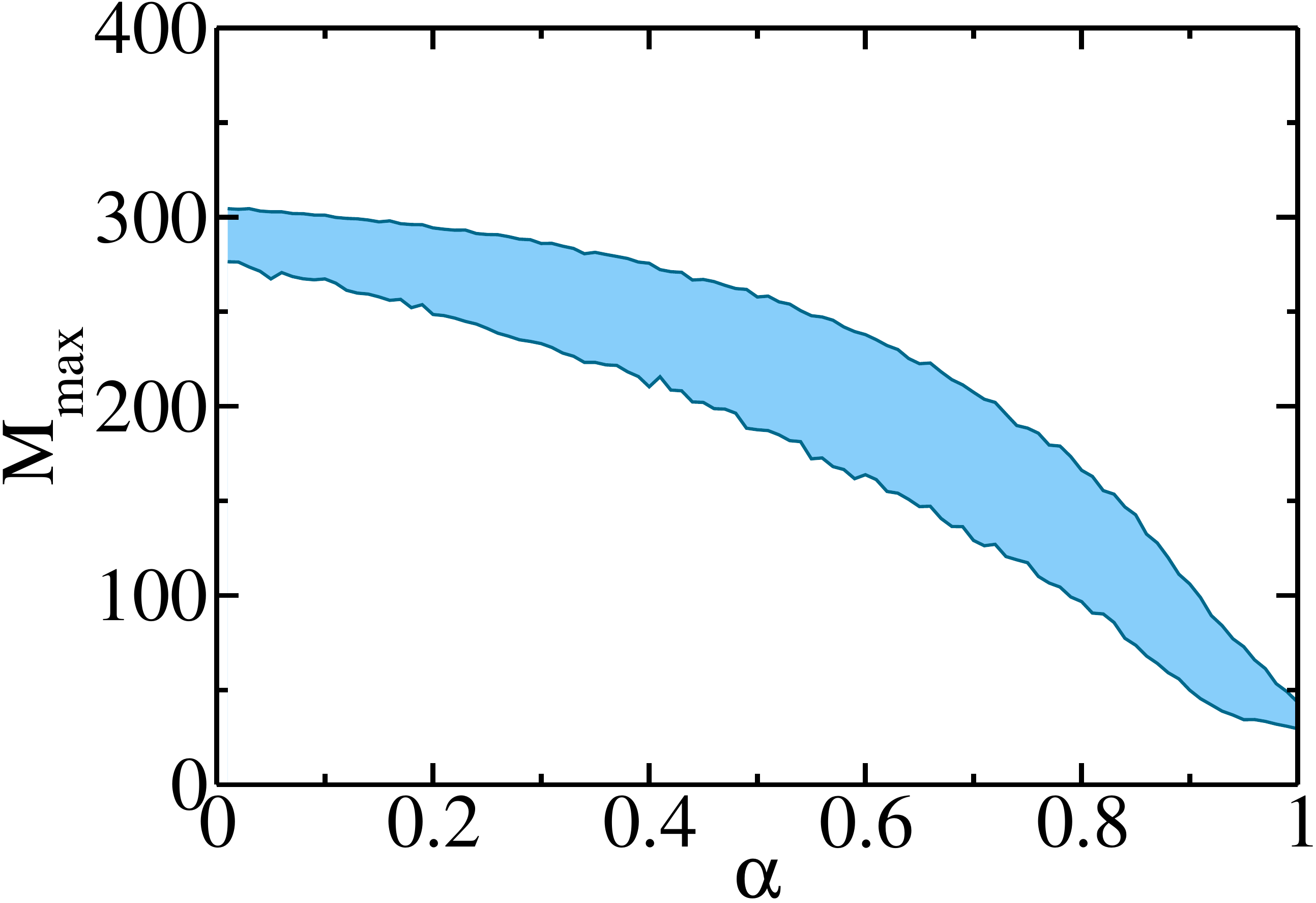}
  \put(65,55){Taipei}
\end{overpic}
\caption{ $M_{max}$ as a function of $\alpha$ using the actual
  temperature and humidity data for the period May 2011-April
  2012. Light blue region: 90\% of the 1000 realizations for the
  synthetic rainfall using the average rainfall and the mean number of
  rainy days per month corresponding to the period 1981-2010 (see
  Appendix ``Weather in Taiwan'').}\label{fig.MAAlph}
\end{figure}

In this section we apply the model of synthetic rainfall time series
to further analyze the effect of rainfall on the mosquito population
in Taiwan. In Fig.~\ref{fig.MAAlph} we show the maximum peak $M_{max}$
of adult mosquito abundance in the summer of 2011 in Taipei and
Kaohsiung; here we use the actual temperature and humidity time series
and the synthetic rainfall for various values of $\alpha$. For
simplicity, we use the same value of $\alpha$ for all months. We
integrate our mosquito model over a period of one year, setting as
initial condition $E_D=E_W=L=P=M=10$ in May 2011, and we use the
average rainfall and the mean number of rainy days per month
corresponding to the period 1981-2010 for the synthetic rainfall
model. From the figure we observe that the peak of abundance is a
decreasing function with the heterogeneity $\alpha$: since the monthly
precipitation is concentrated in fewer days when $\alpha$ increases,
the mosquitoes have often a smaller amount of water available to
reproduce. Similar plots were obtained for Hualien and Taichung (not
shown here). 

\begin{figure}[H]
\begin{overpic}[scale=0.25]{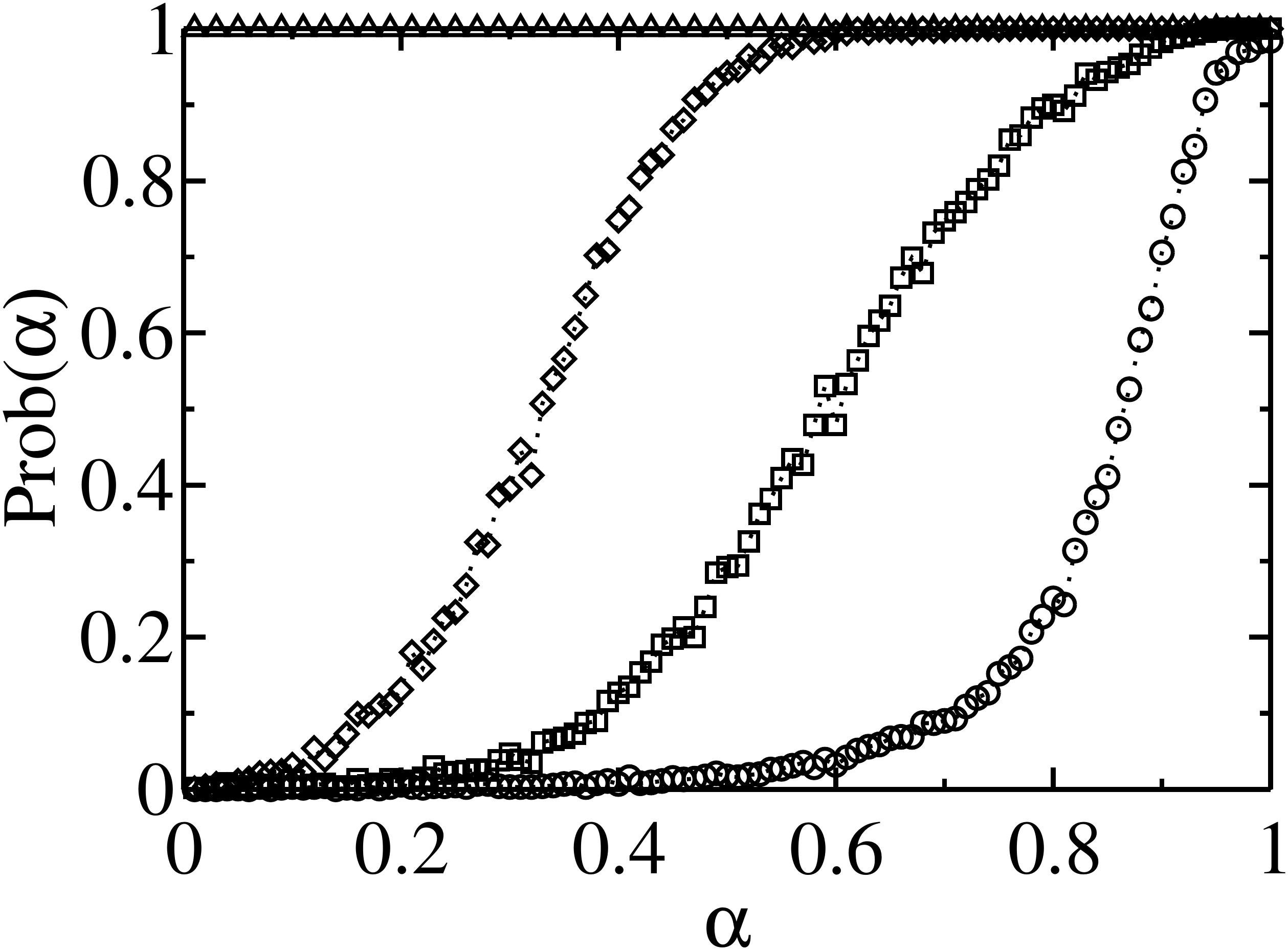}
  \put(15,50){(a)}
\end{overpic}
\begin{overpic}[scale=0.25]{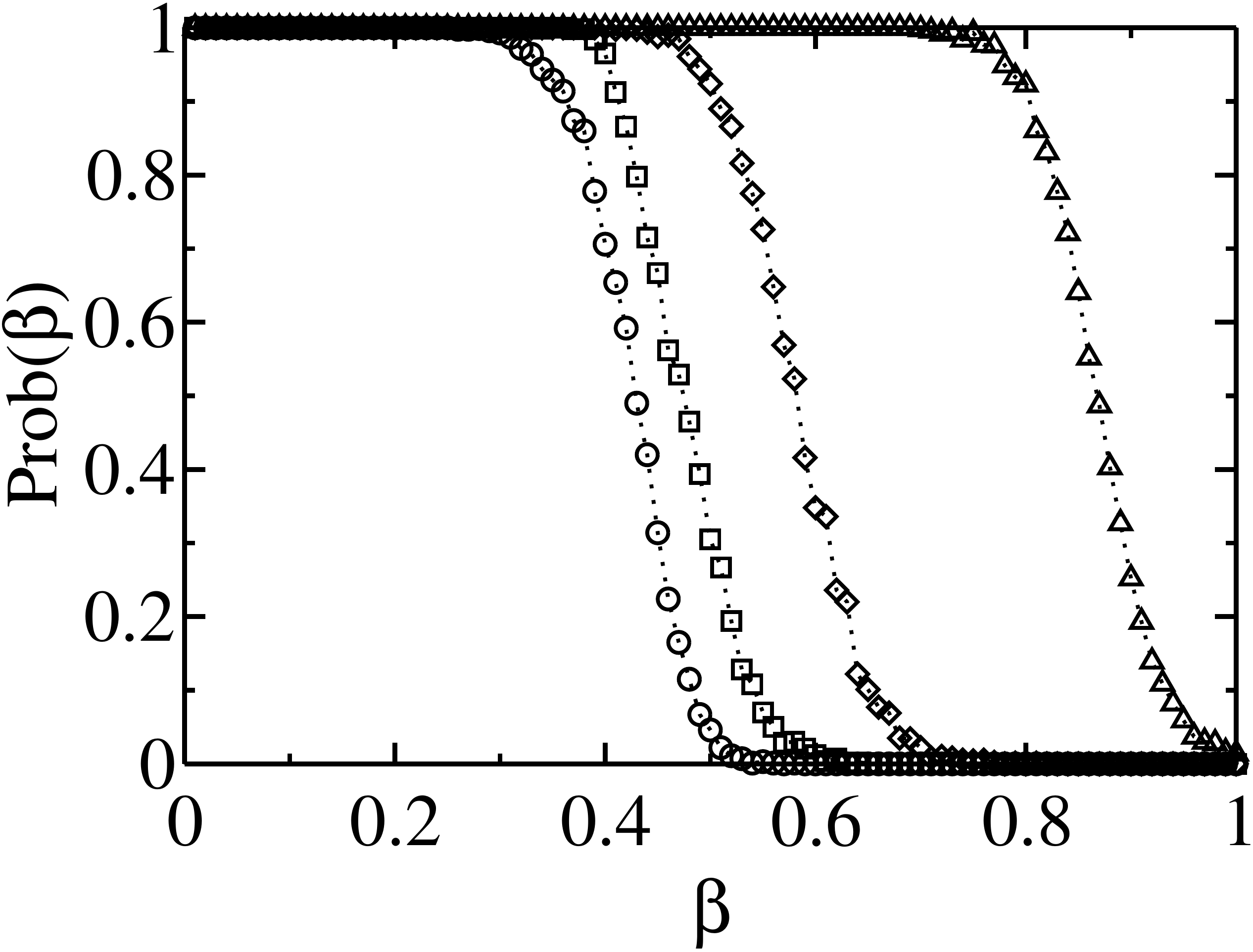}
  \put(15,50){(b)}
\end{overpic}
\caption{{\it Aedes aegypti} extinction probabilities in Kaohsiung
  (circles), Hualien (squares), Taichung (diamonds), and Taipei
  (triangles) as (a) functions of $\alpha$ using the calibrated values
  of Table~\ref{tab.matAjus}; and (b) as functions of $\beta$ with
  $\alpha=0.45$. Dotted lines serve as a guide to the eye. The
  results are averaged over 1000 realizations of the synthetic
  rainfall model. }\label{fig.gambet}
\end{figure}

We also observe that the four cities in Taiwan have favorable
conditions for mosquito breeding during the summer. An important
question is whether the mosquito population that invades an unoccupied
area in spring-summer is able to survive the winter season, at least
as dried eggs. To test this, we integrate the mosquito abundance
equations over the period May 2011- April 2012 for the four cities and
compute the {\it Ae. aegypti} extinction probability $Prob(\alpha)$ at
the end of April 2012 for different rainfall heterogeneity values
$\alpha$. Note that we consider that a mosquito species is extinct if
the number of members in each compartment is null. In
Fig.~\ref{fig.gambet}a we observe that $Prob(\alpha)$ is an increasing
function with $\alpha$ for all the cities, which is consistent with a
decreasing of $M_{max}$ with $\alpha$ as it was shown in
Fig.~\ref{fig.MAAlph}. Additionally, we plot in Fig.~\ref{fig.gambet}b
the extinction probability for different values of the oviposition
rate $\beta$ with $\alpha=0.45$. In this figure we observe that
Kaohsiung is the most favorable city for the establishment of
mosquitoes. In turn, we note that the probability $Prob(\beta)$ is
vanishingly small for $\beta\gtrsim 0.50\equiv\beta_c $ in Kaohsiung,
which suggests that, in order to model {\it Ae. aegypti} populations
in this city with a synthetic rainfall time series, we should use an
oviposition rate $\beta$ above $\beta_c$.

On the other hand, although Taichung is the second city with the
driest winter, our model predicts that the second most favorable city
for {\it Ae. aegypti} after Kaohsiung is Hualien, which has slightly
warmer winters than Taichung. This suggests that temperature can
compensate the negative effect of winter rainfall.

\subsubsection{Why is {\it Ae. aegypti} not endemic to Hualien?}

It is interesting to further discuss the reasons why {\it Ae. aegypti}
is not endemic to Hualien. Having a temperature similar to Kaohsiung
and total winter precipitations that are about half those in Taipei
(see Appendix Sec. ``Weather in Taiwan''), Hualien is the second
most favorable city for the establishment of {\it Ae. aegypti}. From
Fig.~\ref{fig.gambet} we note that for $0.50 \lesssim \beta \lesssim
0.60$ it would be possible to reproduce the real geographic
distribution of {\it Ae. aegypti} in Taiwan. However, if the true
value of the oviposition rate $\beta$ is slightly greater than 0.6,
the fact that there are no mosquitoes in Hualien may be due to
variables we did not consider in our model, such as competition with
other mosquito species~\cite{reiskind2009effects} or local
predators. In addition, Han and Chuang~\cite{han1981air} suggested
that Hualien presents the highest amount of fungi concentration in the
whole country, due to Asian dust storms. For instance, fungal traps in
this city revealed the presence of Periconia and Torula (see
Ref.~\cite{ho2005characteristics}). Notably, a recent study found that
similar fungi were the cause of mortality of {\it Ae. aegypti} eggs in
Chaco, Argentina (see Ref.~\cite{gimenez2015cold}).

\subsubsection{Extinctions and dry season length}

Another relevant aspect of a particular rainfall regime is how the
duration of the dry season affects the establishment and survival of
{\it Ae.aegypti} in a city. To study this effect, we will use a
model of synthetic rainfall time series in which the amount of monthly
rainfall $ \mathscr{P}(m)$ and the number of rainy days
$\mathscr{D}(m)$ are given by the following functions,

\begin{eqnarray}
  \mathscr{P}(m)&=& (P_{max}-P_{min})\left(\frac{1}{2}+\frac{1}{2}\cos\left(\frac{2\pi}{12}(m-m_0)\right)\right)^{\gamma}+\nonumber\\
  &&+P_{min},\label{eq.SecLL1}\\
  \mathscr{D}(m)&=& (D_{max}-D_{min})\left(\frac{1}{2}+\frac{1}{2}\cos\left(\frac{2\pi}{12}(m-m_0)\right)\right)^{\gamma}+\nonumber\\
  &&+D_{min},\label{eq.SecLL2}
\end{eqnarray}
where $\gamma \in [0,\infty)$ controls the duration of the dry season
  (the higher the $\gamma$, the longer the duration of the dry
  season); $m_0$ is the rainiest month of the year, $P_{max}$ is the
  total precipitation in the rainiest month (in millimeters), and
  $P_{min}$ that in the driest month. The variables $ D_{max}$ and
  $D_{min}$ have similar interpretations for the number of rainy
  days. Note that $\mathscr{P}(m_0)=P_{max}$ and
  $\mathscr{D}(m_0)=D_{max}$ for all values of $\gamma$. Analogously,
  $\mathscr{P}(m_0-6)=P_{min}$ and $\mathscr{D}(m_0-6)=D_{min}$ for
  all values of $\gamma$. Here, we set $m_0=7$ (July). Since the
  number of rainy days is an integer number but in
  Eq.~(\ref{eq.SecLL2}) $\mathscr{D}(m)$ is a real number, for each
  rainfall time series realization we choose the number of rainy days
  to be [$\mathscr{D}(m)$] and [$\mathscr{D}(m)$]+1 with probabilities
  $\mathscr{D}(m)$-[$\mathscr{D}(m)$] and
  $1-(\mathscr{D}(m)-[\mathscr{D}(m)])$, respectively (here $[\cdots]$
  stands for the integer part).

\begin{figure}[H]
\centering
\begin{overpic}[scale=0.50]{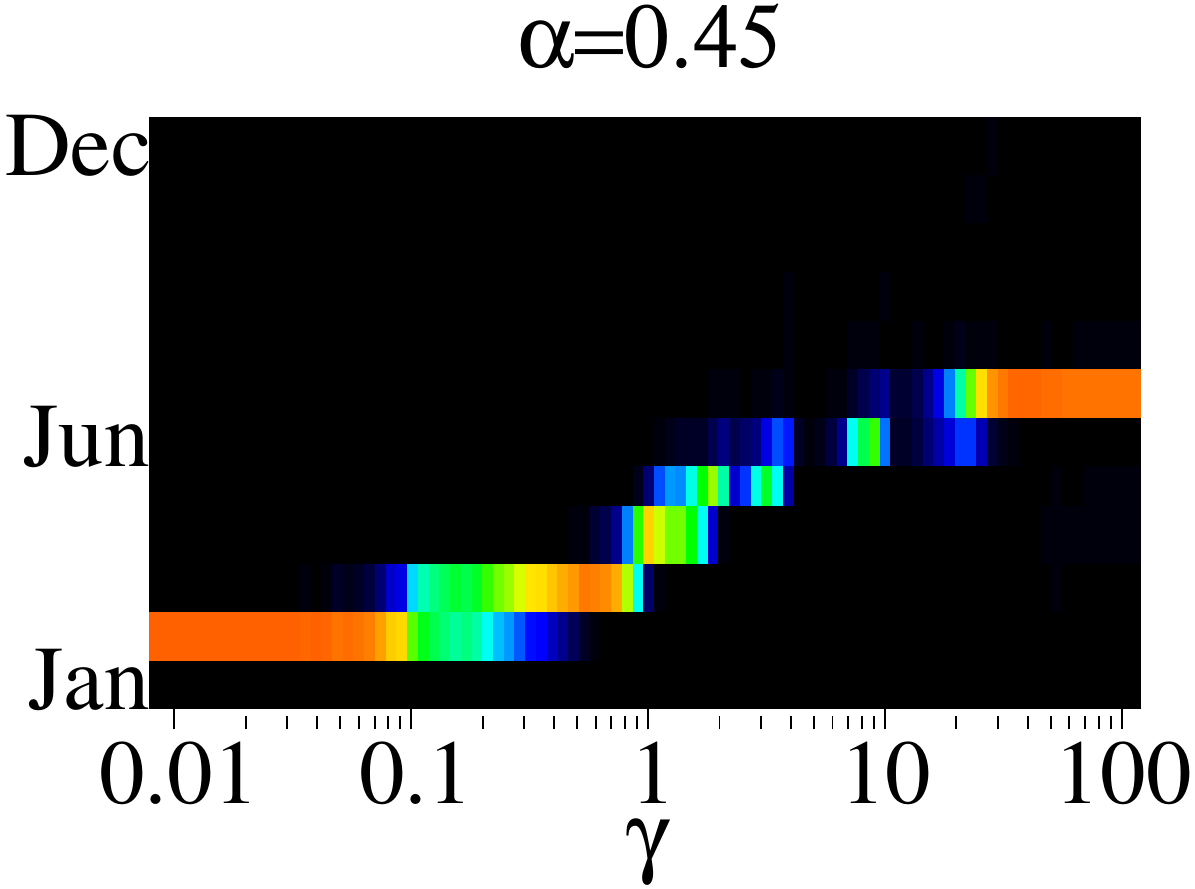}
  \put(1,80){(a)}
\end{overpic}
\begin{overpic}[scale=0.35]{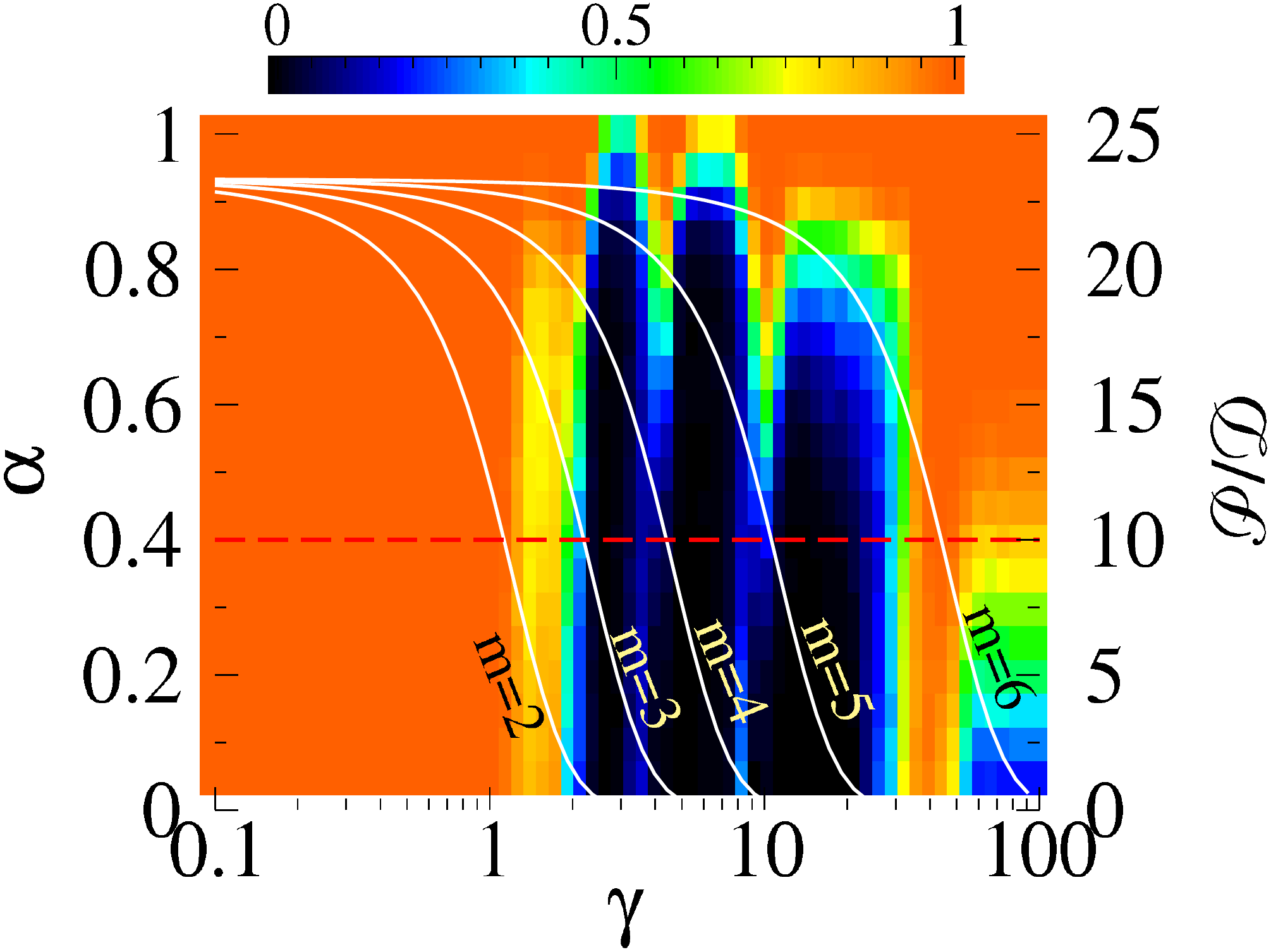}
  \put(1,68){(b)}
\end{overpic}
\caption{Effect of the dry period on the extinction probability. Using
  the model of synthetic rainfall time series, we compute (a) the
  probability of extinction of mosquitoes for each month in Taipei as
  a function of $\gamma$ for $\alpha=0.45$, and (b) the cumulative
  probability of extinction of {\it Ae. aegypti} in Taipei for
  different values of $\alpha$ and $\gamma$. We compute the cumulative
  probability of extinction after the second year from the date of the
  initial condition. Panel (b) also shows the average amount of
  precipitation in a rainy day $\mathscr{P}(m)/\mathscr{D}(m)$ (solid
  white lines) for $m=2,3,4,5,6$ and the rainfall threshold
  $R_{thres}=10$[mm] (see Eq.~(\ref{eq.hill}) and dashed red
  line). The monthly rainfall distribution and the number of rainy
  days follow Eqs.~(\ref{eq.SecLL1}) and~(\ref{eq.SecLL2}). We set:
  $P_{max}=350$, $P_{min}=0$, $D_{max}=15$, $D_{min}=1$, and
  $m_0=7$. The initial condition is $E_D=E_W=L=P=M=10$ on May 1, 2011
  and we model the dynamics of mosquito abundance for four years. Cold
  colors represent a low probability of extinction (black corresponds
  to a null probability) and warm colors a probability close to 1. The
  results are averages obtained over 1000 realizations of rainfall time
  series.}\label{fig.Mgam}
\end{figure}

In Fig.~\ref{fig.Mgam}a we show the probability of extinction of {\it
  Ae. aegypti} for each month as a function of $\gamma$, for
$\alpha=0.45$. In Appendix Sec. ``Effect of the dry season on
{\it Ae.aegypti}'' we show the same plot for $\alpha=0.10$ and $0.80$;
in all cases we use the actual temperature and humidity time series of
Taipei. We observe that for low values of $\gamma$, that is to say,
when the amount of precipitation is high in almost every month, the
extinction occurs in winter. This is consistent with the Tsuda and
Takagi's hypothesis, which states that winter rainfalls could cause
eggs to hatch in a period with low temperatures (see the
Introduction). As $\gamma$ increases, the maximum probability of
extinction moves forward in time: dry periods are longer and more
intense. Finally for the highest values of $\gamma$, the extinction
occurs in summer: since the (extremely) dry season lasts almost 11
months, very few viable eggs survive until the (short) wet season;
these cannot sustain a mosquito population after a rainfall
event. Note that extinctions never occur in the fall.

In Fig.~\ref{fig.Mgam}b we show the effect of the daily rainfall
heterogeneity ($\alpha$) and the duration of the dry season ($\gamma$)
on the cumulative probability of extinction of {\it Ae. aegypti} in
the city of Taipei. Again, we use the actual temperature and humidity
time series of Taipei. We obtain, as expected, that for very low
values of $\gamma$ mosquitoes become rapidly extinct since rainfalls
are abundant throughout the year. It should be noted that the
probability of extinction does not depend on the daily rainfall
variability ($\alpha$). Therefore, when there is no dry season, the
amount of {\it monthly} rainfall (in millimeters) is a sufficient
predictor for determining the absence of mosquitoes. However, as
$\gamma$ increases, we observe that the probability of extinction
drops, and hence in this parameter domain we can consider that Taipei
resembles Kaohsiung. The dry winter allows a bigger egg population to
survive; when they hatch in the rainy season the temperature is
already high enough to allow these eggs to prosper. Additionally, in
this case the probability of extinction also depends on $\alpha$. This
suggests that in areas with dry winter months but without a permanent
{\it Ae. aegypti} population, the role of the variability in the daily
precipitation as an explanatory variable for these extinction cases
should be further explored.

On the other hand, note that as $\gamma$ grows for a fixed value of
$\alpha$, the extinction probability increases and decreases several
times. In Fig.~\ref{fig.Mgam}b we show that local transitions from
high to low extinction probabilities take place when the average
amount of precipitation in a rainy day on month $m$, $i.e.$
$\mathscr{P}(m)/\mathscr{D}(m)$ is approximately equal to the rainfall
threshold (see Eq.~(\ref{eq.hill})). An extinction may occur, for
instance, in a month when rainfalls can flood a certain quantity of
eggs but the amount of available water is not enough to sustain a
larval population; as $\gamma$ increases this month will become a
``dry'' month, so the probability of extinction diminishes because
fewer eggs hatch under detrimental conditions. Finally, for very high
values of $\gamma$, the extinction probability again increases, which
is consistent with the results shown in Fig.~\ref{fig.Mgam}a for a
very dry region.

\section{Summary}\label{sec.SumFW}
Since {\it Ae. aegypti} is expanding its territory, potentially
contributing to the spread of diseases such as Dengue, Yellow Fever, Chikungunya and
Zika \cite{gubler2004changing,pialoux2007chikungunya,zhang2017spread}
to new areas, it is important to assess how different climatic
variables can shape the present and future geographic range of this
mosquito. We have developed a model for the {\it Ae. aegypti}
population that takes into account the susceptibility of its immature
stages to winter rains. We hypothesize that locations with cool, rainy
winters are inimical to this species and test the model by applying
it to four cities in Taiwan, reproducing the observed presence (or
absence) of {\it Ae. aegypti}. In particular, we find
that the reason {\it Ae. aegypti} is endemic to Kaohsiung is that it
is the city with the lowest precipitation in winter. We also
introduce a procedure to generate rainfall time series to explore the
effect of different rainfall regimes.

Applying our rainfall model to Taiwan, we find that as the
precipitation heterogeneity ($\alpha$) increases, the peak of mosquito
abundance during the summer decreases. On the other hand, a reduction
in the heterogeneity of daily rainfall decreases the probability of
extinction in all the cities, except Taipei. We also obtain that the
presence or absence of {\it Ae. aegypti} depends on a delicate balance
between different climatic variables in the regions near the
$10^{\circ}$C winter isotherms.

Finally, we studied the effect of dry season duration on
mosquito survival and found that, as it increases, the survival
probability also increases. However, because eggs become gradually
nonviable, for very long droughts, the likelihood of
extinction rises again. 

Given that climate change is likely to make the world wetter,
predicting the evolution of the geographical habitat of {\it
  Ae. aegypti} is not easy: while an increase in the temperature
should shift its boundaries towards higher latitudes, an increase in
the winter rainfall would be detrimental to its enlargement. The model
presented here can be useful to address this challenge.

\section*{Acknowledgments}
This work was supported by SECyT-UNC (projects 103/15 and 313/16) and
CONICET (PIP 11220150100644). We also thank Mgter. Laura L\'opez for
useful discussions.

\setcounter{equation}{0}
\setcounter{table}{0}
\setcounter{figure}{0}
\renewcommand{\theequation}{S\arabic{equation}}
\renewcommand{\thetable}{S\arabic{table}}
\renewcommand\thefigure{S\arabic{figure}} 
\section{Appendix}

\subsection{Weather in Taiwan}\label{Si.Weath}

-
\begin{figure}[H]
\begin{minipage}[c][8cm][t]{.42\textwidth}
  \vspace*{\fill}
  \centering
  \includegraphics[width=7.5cm,height=9.5cm]{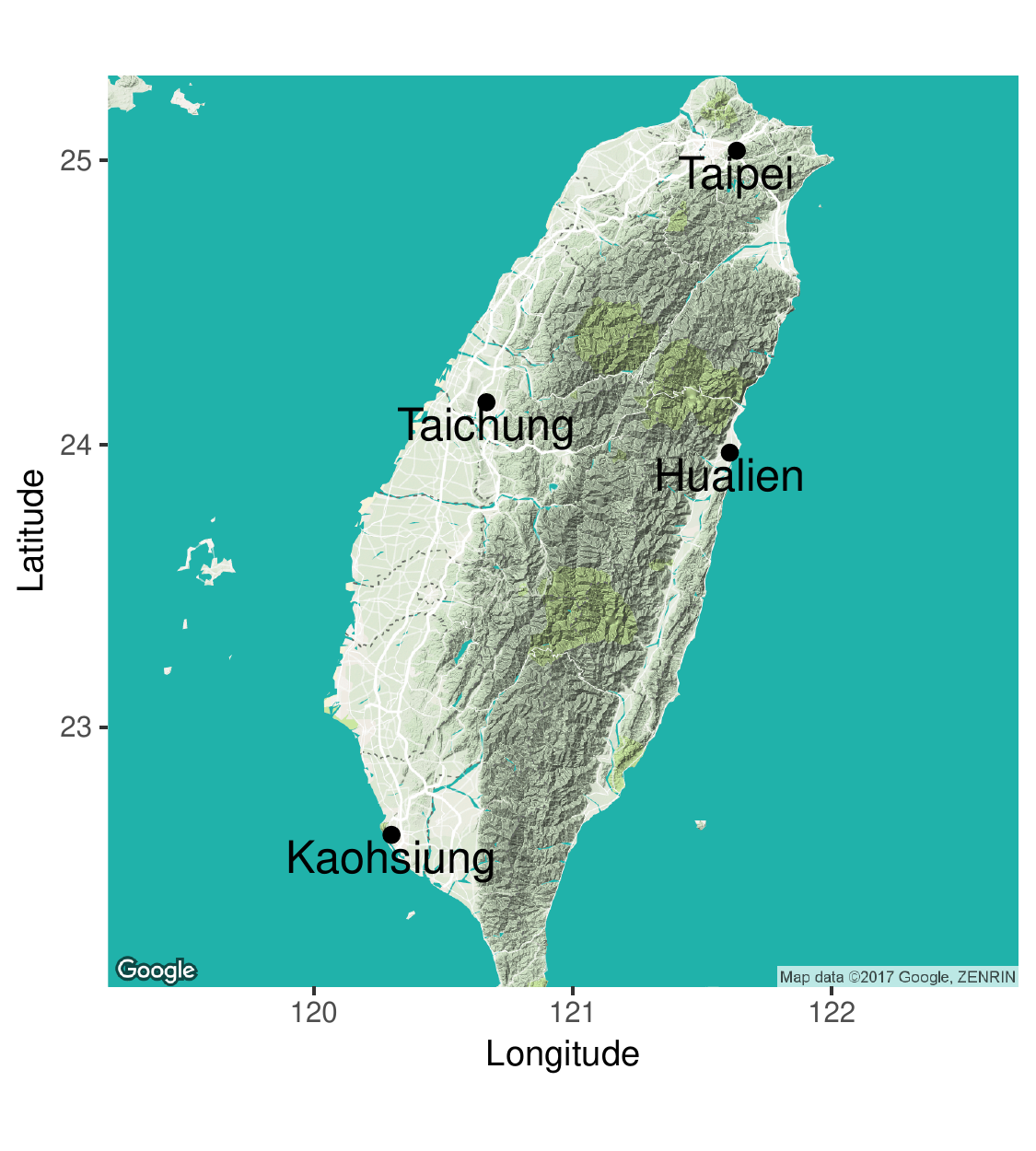}\vspace{-1cm}
\end{minipage}%
\begin{minipage}[c][8cm][t]{.42\textwidth}
  \centering
  \vspace{0.7cm}
  \includegraphics[scale=0.10]{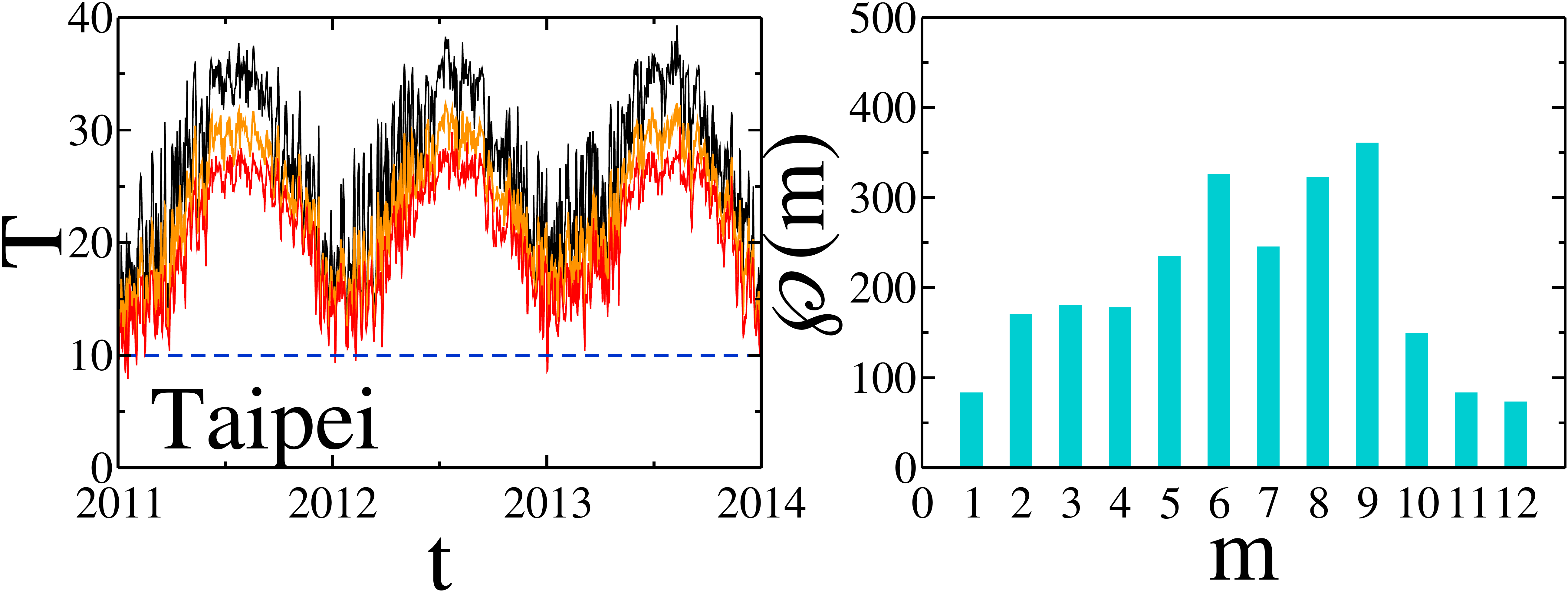}
  \includegraphics[scale=0.10]{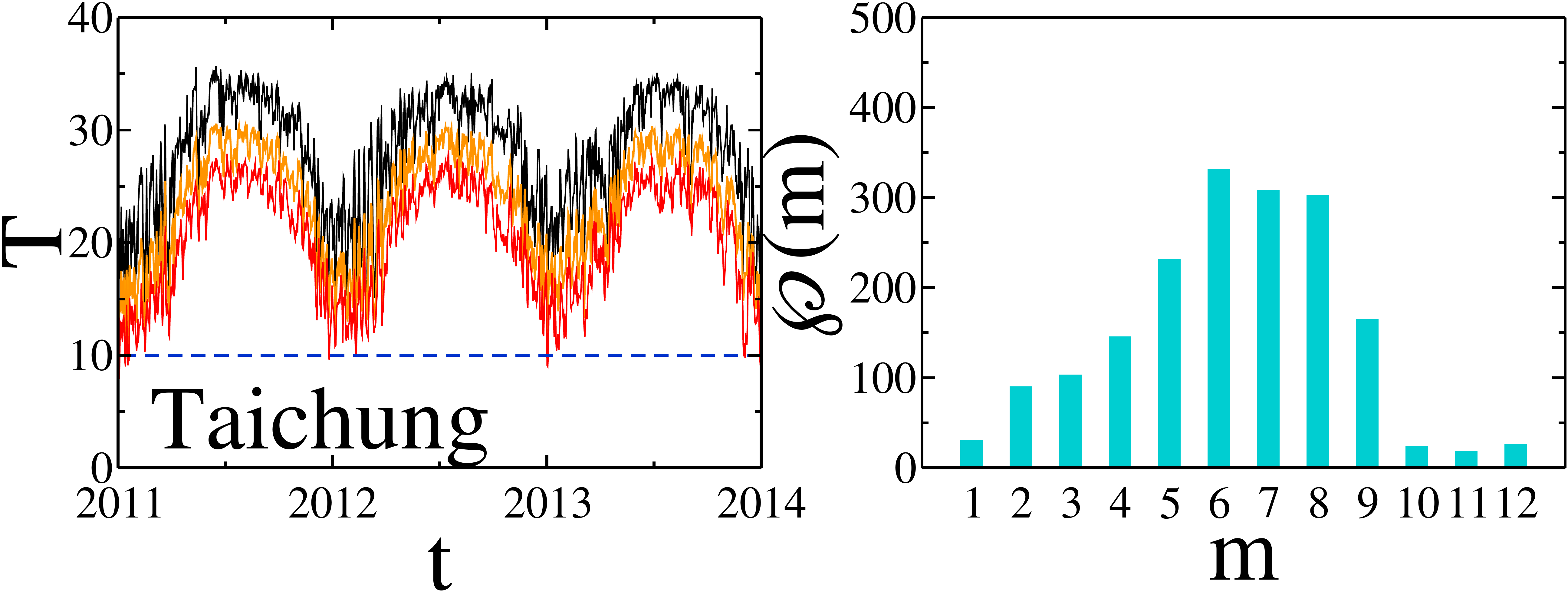}
  \includegraphics[scale=0.10]{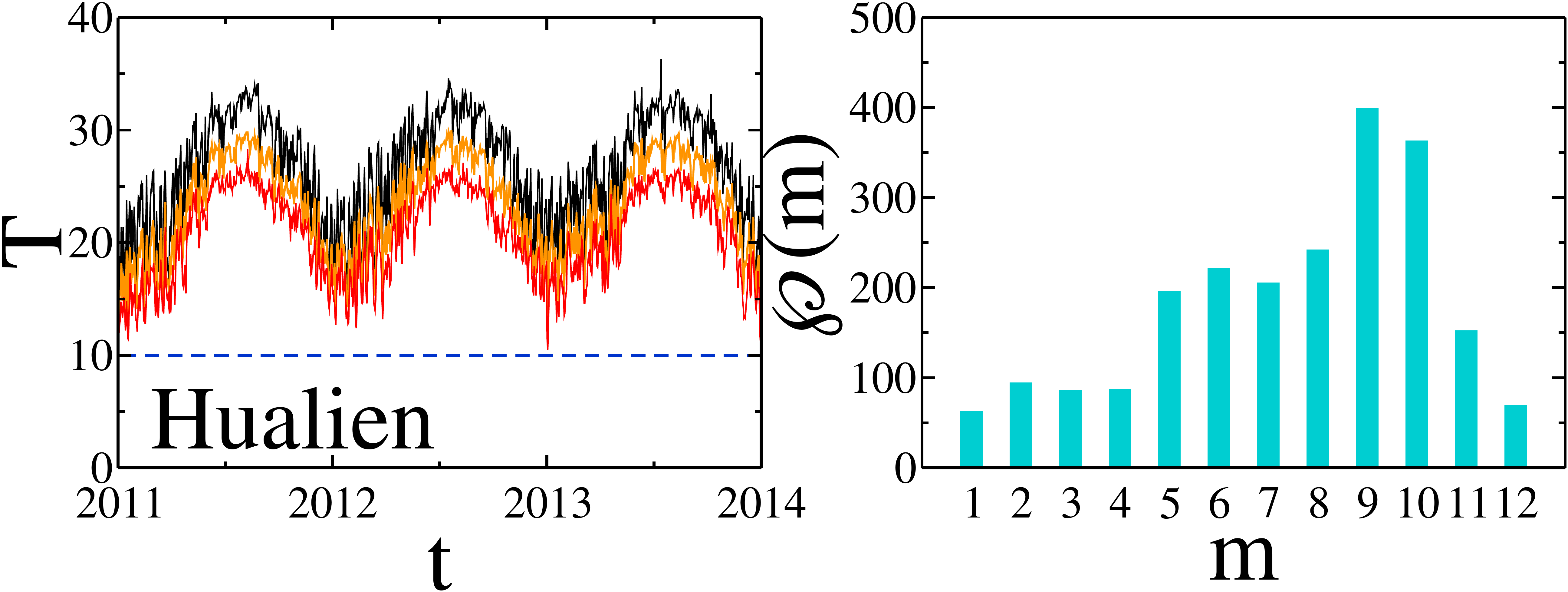}
  \includegraphics[scale=0.10]{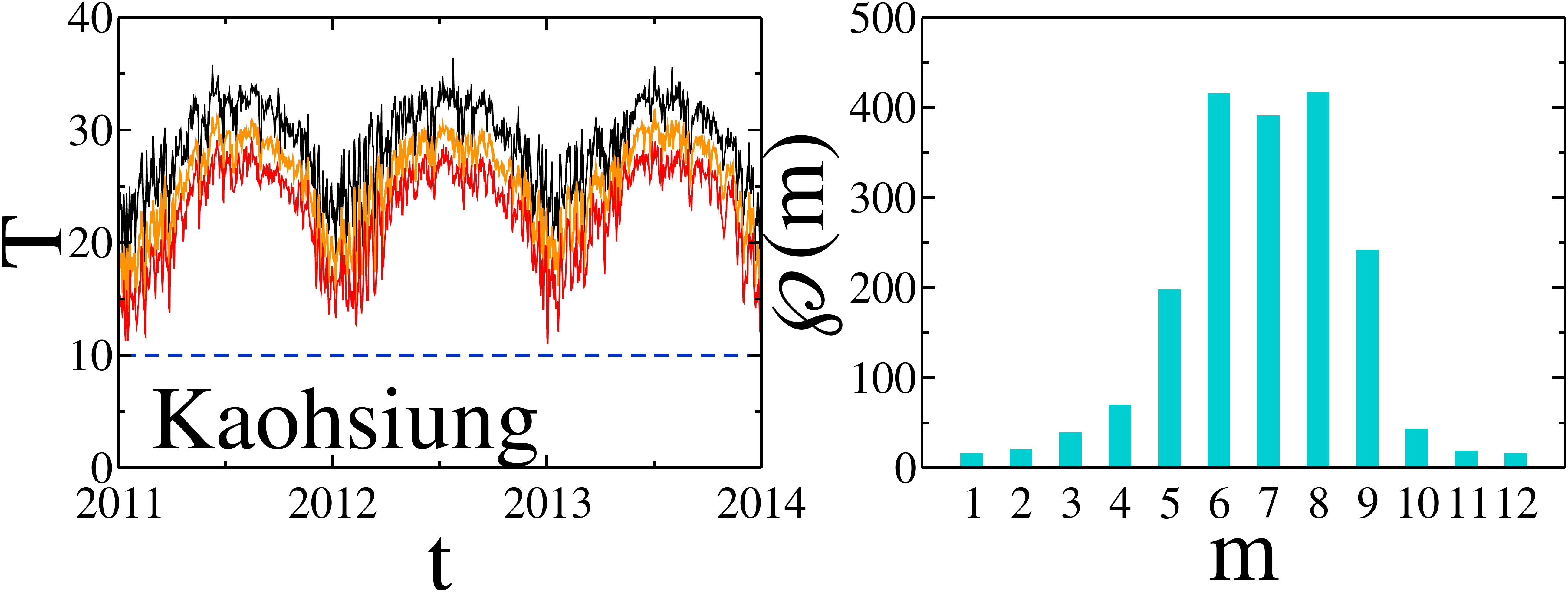}
\end{minipage}
\vspace{1cm}
\caption{Left panel: Map showing the four cities of Taiwan for which
  we apply our mosquito model using a synthetic rainfall time series:
  Taipei, Taichung, Hualien, and Kaohsiung~\cite{Exxon_03}. Center
  panels: Maximum (black line), average (orange line) and minimum
  (red line) temperatures for each city in the period 2011-2013. The horizontal
  dashed line stands for the $10^{\circ}$C isotherm. Rightmost panels:
  Average monthly rainfall $\mathscr{P}(m)$ in the period 1981-2010 (bar
  plots), with $m=1$ for January and $m=12$ for December.}\label{fig.DatTaiw}
\end{figure}

\subsection{Effect of the dry season on {\it Ae.aegypti}}

-

\begin{figure}[H]
\centering
\begin{overpic}[scale=0.50]{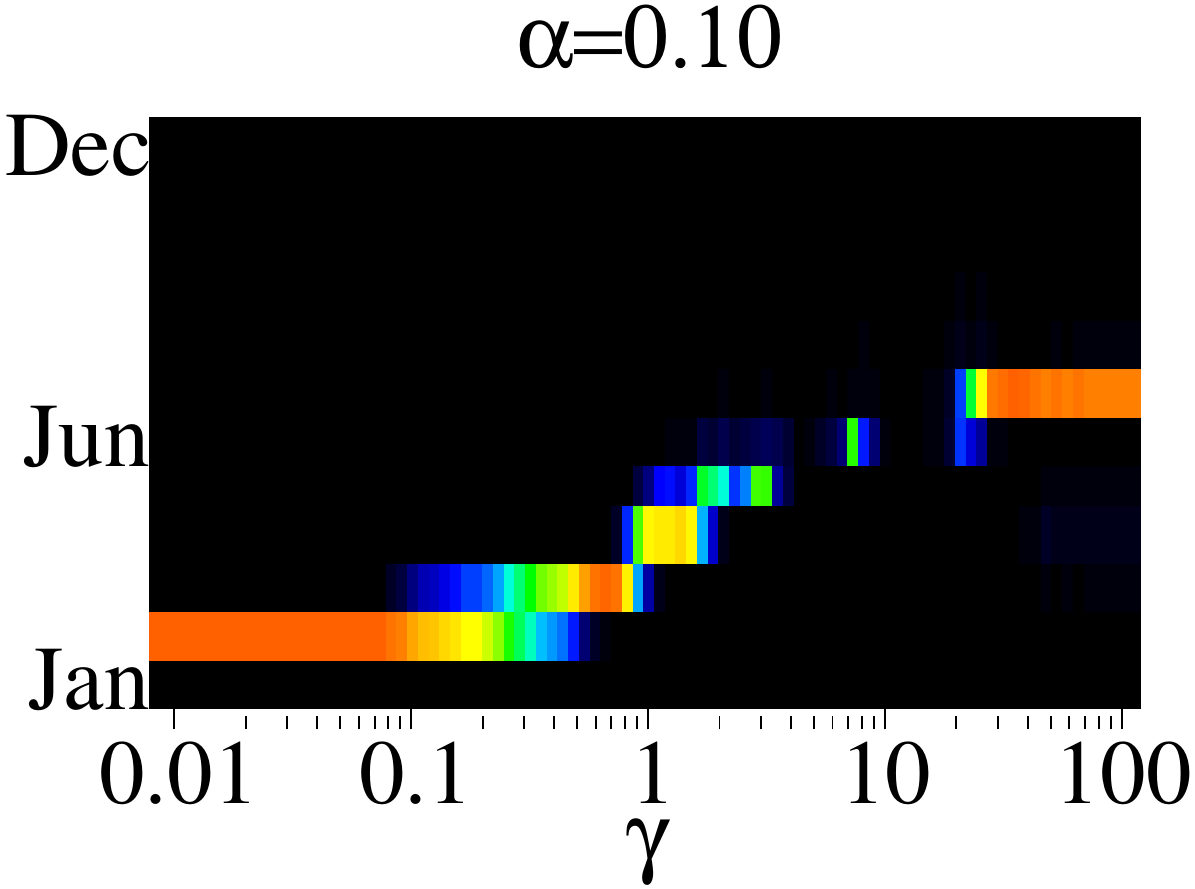}
  \put(1,80){(a)}
\end{overpic}
\begin{overpic}[scale=0.50]{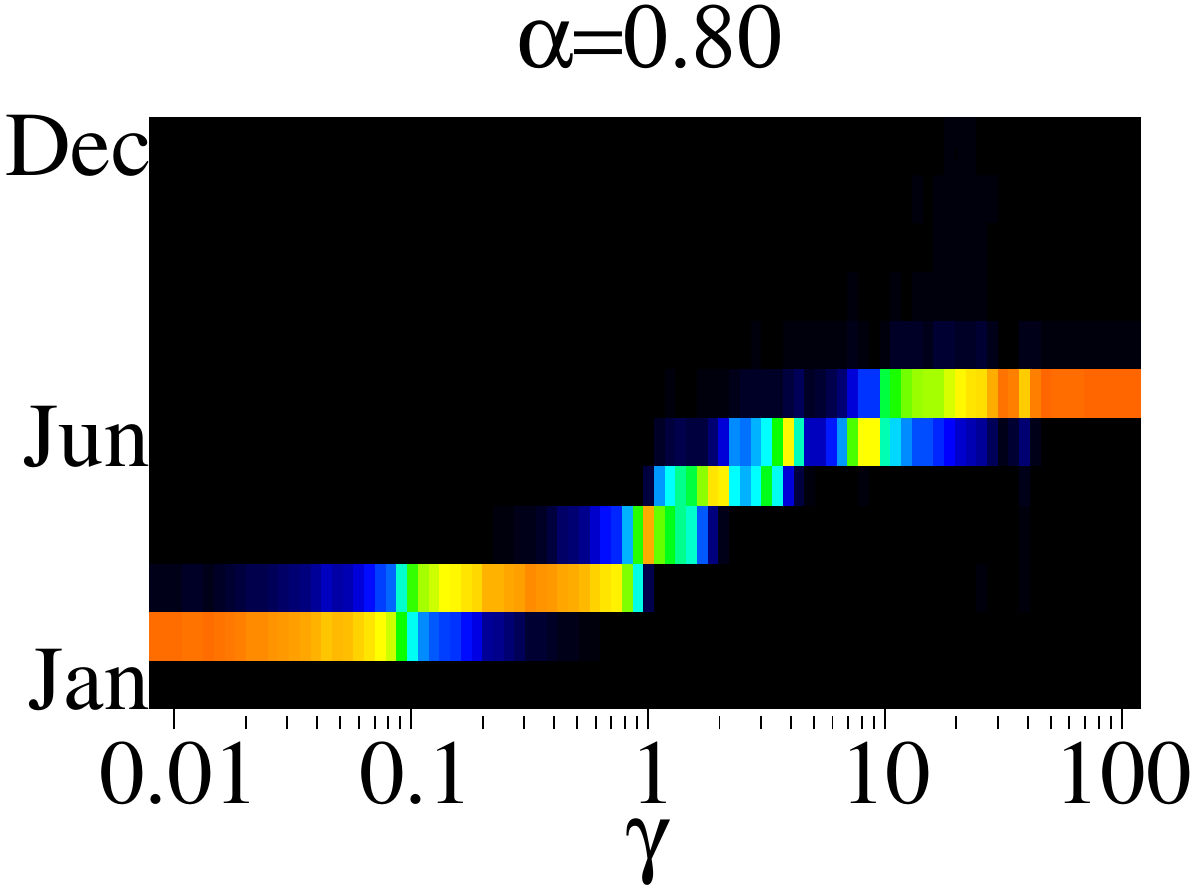}
  \put(1,80){(b)}
\end{overpic}
\caption{Probability of extinction of mosquitoes for each month in
  Taipei as a function of $\gamma$, using the model of synthetic
  rainfall time series for two $\alpha$ values: 0.10 and 0.80. The
  monthly rainfall distribution and the number of rainy days follow
  Eqs.~(\ref{eq.SecLL1}) and~(\ref{eq.SecLL2}). We chose:
  $P_{max}=350$, $P_{min}=0$, $D_{max}=15$, $D_{min}=1$, and
  $m_0=7$. The initial condition is $E_D=E_W=L=P=M=10$ on May 1,
  2011. We model the dynamics of mosquito abundance for four
  years. Cold colors represent a low probability of extinction (black
  corresponds to a null probability) and warm colors a probability
  close to 1. The results were averaged over 1000 realizations of
  rainfall time series.}
\end{figure}

\begin{figure}[H]
\centering
\begin{overpic}[scale=0.25]{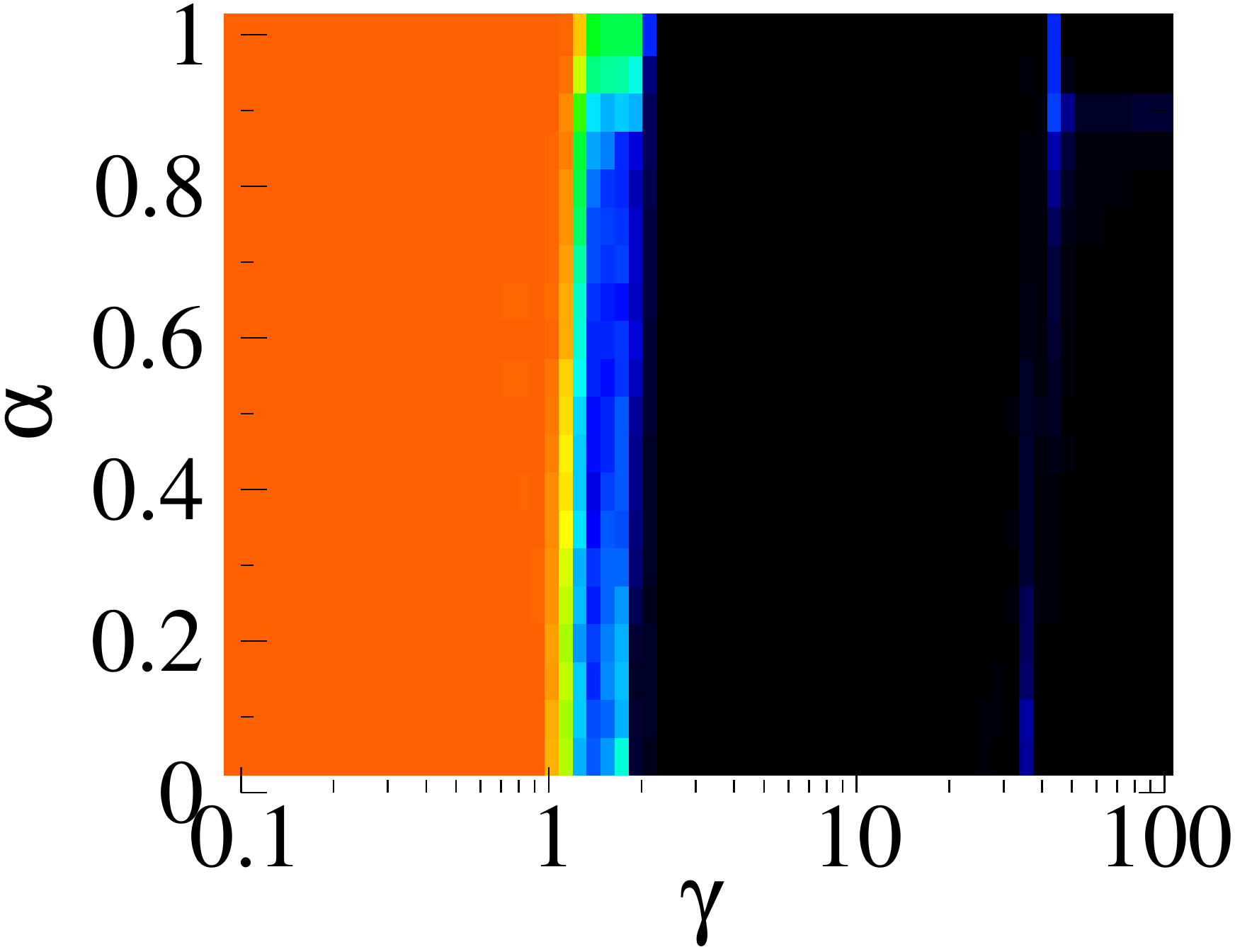}
  \put(1,68){(a)}
\end{overpic}
\begin{overpic}[scale=0.25]{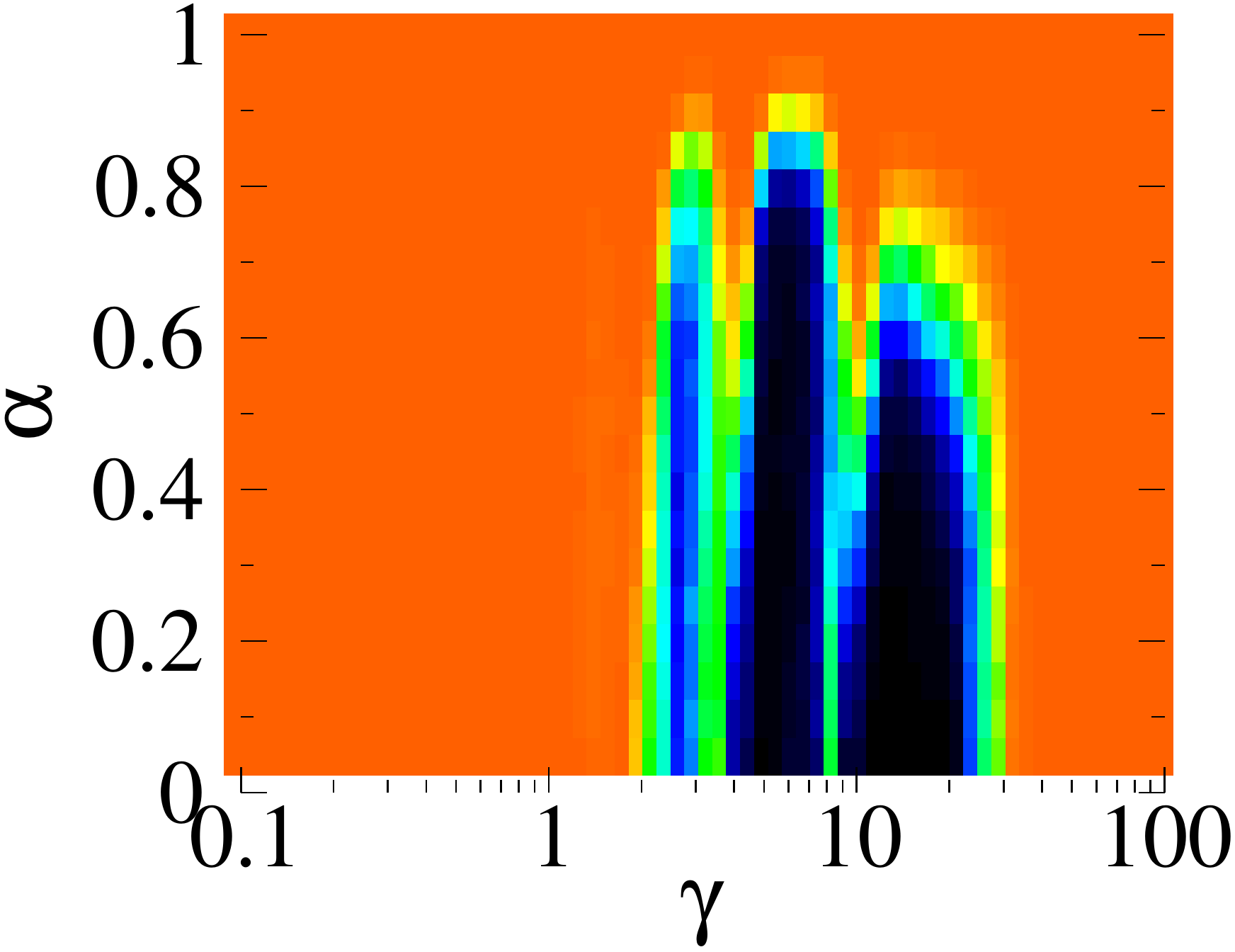}
  \put(1,68){(b)}
\end{overpic}
\begin{overpic}[scale=0.25]{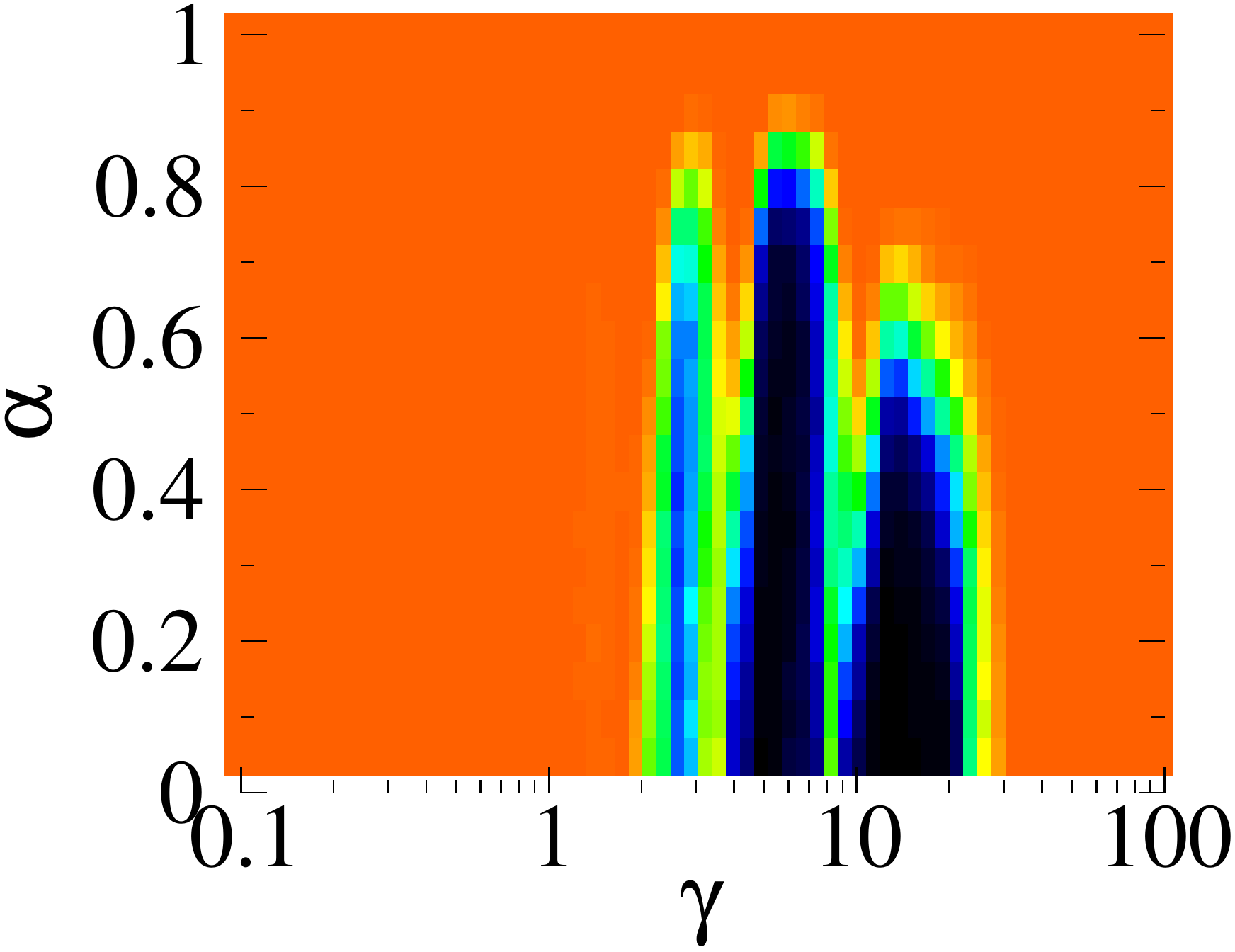}
  \put(1,68){(c)}
\end{overpic}
\caption{ Cumulative probability of extinction of {\it Ae. aegypti} in
  Taipei for different values of $\alpha$ and $\gamma$, using the
  model of synthetic rainfall time series. The monthly rainfall
  distribution and the number of rainy days follow
  Eqs.~(\ref{eq.SecLL1}) and~(\ref{eq.SecLL2}). We chose:
  $P_{max}=350$, $P_{min}=0$, $D_{max}=15$, $D_{min}=1$, and
  $m_0=7$. The initial condition is $E_D=E_W=L=P=M=10$ on May 1,
  2011. We compute the cumulative probability of extinction after one
   (a), three (b) and four (c) years. The colors have the same
  meaning as in Fig.~\ref{fig.Mgam}. }
\end{figure}

\subsection{Model calibration}\label{ap.Calib}
For the calibration of our mosquito population model, we use a Latin
hypercube sampling method for the parameters $\beta$, $k$, $H_{max}$,
and $ K_{max}$, and compute the mean square distance between the
predicted abundance of mosquitoes and the actual weekly adult mosquito
abundance data of Kaohsiung in the period January 2010 - December 2012
(obtained from Ref.~\cite{Exxon_04}). Note that we re-scaled the
actual abundance data, so it is about 100 mosquitoes per block in the
summer. We set as initial conditions, $E_D=E_W=L=P=M=100$. For the
same set of points, we repeat the process for Taipei, Taichung, and
Hualien~\footnote{For the calibration in Taipei and Kaohsiung, we
  integrate our equations over the period January 2010- December
  2012. For Hualien and Taichung we adjust our model in the period
  January 2011- December 2012, since the data of temperature and
  humidity in 2009 are not available.}. In all the cases, we start the
integration six months before the time from which we compute the mean
square distance, in order to minimize the effect of the initial
conditions. Note that we use the actual values of temperature,
humidity, and rainfall for the calibration process. Finally, we choose
the point ($\beta$,$k$,$H_{max}$,$ K_{max}$) that minimizes the sum of
the square distances corresponding to all the cities.

\subsection{Integration of the equations of our model}\label{ap.IntEq}

We model a system of compartmental difference-differential equations
which we integrate numerically using the Euler method with $\Delta
t=0.01$, where $t$ is given in days. If for any time step the
population of a compartment is negative, then we change its value to
0. Since the time series of temperature, humidity and rainfall are on
a daily time scale, all the parameters that depend on these variables
are considered constant between days $d$ and $d + 1$, i.e. for
$t\in[d,d+1)$, where $d$ is a non-negative integer number. These
  parameters are calculated using the values of temperature, humidity,
  and rainfall for that day. In turn, if it rains on day $d$, then at
  the end of each rainy day, i.e. when $t = d + 1-\Delta t$, a
  fraction of dry eggs go wet as it is explained in
  Sec.~``Rainfall and wet eggs''. Similarly, if on day $d$ the minimum
  temperature is below $10^{\circ}$C, then at time $t=d+1-\Delta t$ we
  apply the rules explained in Sec.~``Survival of {\it Ae. aegypti} at
  low temperatures''. Finally, to incorporate the fact that we model
  discrete populations and to be able to explore extinction processes,
  we will impose that if the population of a compartment is less than
  1 at the end of the integration step $t = d-\Delta t$, then we
  change its value to 0.

\subsection{Calibration of $\alpha$}\label{ap.CalAlp}
In order to compute the value of $\alpha$ that best fits the actual
rainfall time series, we propose a procedure based on the
distance minimization between the variance of the actual rainfall time
series and that of the synthetic one. Namely, we propose the following
steps:
\begin{enumerate}
\item We calculate the total precipitation $\mathscr{P}$ (in
  millimeters) and the number of rainy days $\mathscr{D}$ in a period
  of time (for instance, 30 days) and the variance of the actual daily
  rainfall. We exclude non-rainy days from the computation of this
  variance.
\item We use the fracturing process for different values of $\alpha\in
  [0,1]$ using the actual values of the total rainfall and the number
  of rainy days computed in the previous step. Then we measure the
  variance of the synthetic rainfall series. This step is repeated
  $10^5$ times for each value of $\alpha$.
\item We choose the value of $\alpha$ that minimizes the distance
  between the average variance of the synthetic rainfall and the
  variance of the actual rainfall time series.
\end{enumerate}

To ascertain if this procedure allows us to re-obtain a preset value
of $\alpha$ for a synthetic rainfall time series, we generate these
series for different values of $\alpha$ and then we apply the method
described above. In Table~\ref{tab.AjAlpArt} we show that, although our
method slightly underestimates the values of $\alpha$, the new values
are close enough to the preset ones to confirm the consistency of the
procedure.

\begin{table}[H]
\centering
\caption{Estimated values of $\alpha$ for various synthetic rainfall
  time series. We constructed 100 of these series for each preset value
  of $\alpha$ using a total precipitation of $\mathscr{P}=200$ and
  $\mathscr{D}=10$ rainy days in a period of 30
  days. } \label{tab.AjAlpArt}
\begin{tabular}{|c|c|}
\hline
$\alpha$ (preset value) &$\bar{\alpha}$ (estimated value) \\
\hline
 0.2 &0.16 \\
 0.3&0.27\\
 0.4& 0.35 \\
 0.45& 0.42\\
 0.5&0.45 \\
 0.6&0.58 \\
\hline
\end{tabular}
\end{table}
\subsection{Transition rates}\label{Ap.ModelDet}

\subsubsection{The impact of mean daily temperature on oviposition}
We propose that the number of dry eggs laid per unit time is
proportional to: 1) the number of adult mosquitoes, 2) the oviposition
rate $\beta$ in optimal conditions of temperature, and 3) a factor
$\theta \in [0,1]$, which stands for the effect of temperature on
oviposition. In this work, $\theta$ is given by
\medskip
\begin{eqnarray}\label{eq.theta}
\theta(t)=\left\{%
\begin{array}{ll}
 0.1137 (-5.4+1.8T-0.2124T^2+0.01015 T^3-0.0001515 T^4 ) & \;\;\;\; \text{if}\;\;\;\;11.7<T(t)<37.2\\
0  &\;\;\;\; \text{otherwise},
\end{array}%
\right.
\end{eqnarray}
Here the relation between $\theta$ and the mean daily temperature $T$
was obtained by Yang et al. on the basis of temperature laboratory
experiments (see Refs.~\cite{yang2009assessing,lourencco20142012}), and the
prefactor 0.1137 normalizes $\theta$.

\subsubsection{Maturation rates}
In table~\ref{tab.mat1} we define the maturation rates $m_X$ (where
$X=E$, $L$, and $P$ stand for eggs, larvae, and pupae,
respectively). Equation~(\ref{eq.mature}) defines their relation with
the mean daily temperature.
\begin{table}[H]
\centering
\caption{Notation and definitions of the maturation rate coefficients}
\label{tab.mat1}
\begin{tabular}{|c|p{6.5cm}|c|}
\hline
Quantity & Definition & Value  \\
\hline
$m_E$ & rate at which eggs develop into larvae (days$^{-1}$) & Eq.~(\ref{eq.mature})\\
$m_L$ & rate at which larvae develop into pupae (days$^{-1}$) & Eq.~(\ref{eq.mature})\\
$m_P$ & rate at which pupae develop into mosquitoes (days$^{-1}$) &Eq.~(\ref{eq.mature})\\
\hline
\end{tabular}
\end{table}

\begin{eqnarray}\label{eq.mature}
m_X=R_X \frac{T+T_0}{298}\frac{\exp\left[\frac{\Delta H_A}{R_0}\left(\frac{1}{298}-\frac{1}{T+T_0}\right)\right]}{1+\exp\left[\frac{\Delta H_H}{R_0}\left(\frac{1}{T_{1/2}}-\frac{1}{T+T0}\right)\right]}
\end{eqnarray}
(see Ref.~\cite{barmak2014modelling} and references therein), where $T$ is
the mean temperature (in Celsius), $T_0=273.15^{\circ}$C, and $R_0$ is
the universal gas constant. The other parameters ($R_X$, $\Delta
H_A$, $\Delta H_H$, $T_{1/2}$) are shown in Table~\ref{tab.mat2} for
$X=E$, $L$, and $P$.
\begin{table}[H]
\centering
\caption{Values of $R_X$, $\Delta H_A$, $\Delta H_H$, and $T_{1/2}$ for
  eggs, larvae and pupae}
\label{tab.mat2}
\begin{tabular}{|c|c|c|c|c|}
\hline
X & $R_X$ &  $\Delta H_A$ & $\Delta H_H$&$T_{1/2}$\\
\hline
$E$& 0.24  & 10798 & 100000 & 14184\\
$L$&  0.2088& 26018 & 55990 & 304.6\\
$P$&   0.384& 14931 & -472379 & 148\\
\hline
\end{tabular}
\end{table}

In addition, we set the maturation rate
coefficient for {\it Ae. aegypti} larvae to be equal to zero for
$T<13.4^{\circ}$C~(see Refs.~\cite{chen1988ecological,chang2007differential}).

\subsubsection{The Gillett effect and the intraspecific larval competition}
Wet eggs hatch at a rate proportional to $m_E$, which depends only
on temperature (see Eq.~(\ref{eq.mature})). However, several
studies~\cite{gillett1955variation,gillett1977erratic} suggested
that the hatching process is delayed when the larval population
increases, which is called the Gillett effect. In order to
introduce this effect in our model, we multiply the rate $m_E$ by
\medskip
\begin{eqnarray}\label{eq.CG}
C_G=\left\{%
\begin{array}{ll}
 1-\frac{L}{K_L} & \;\;\;\; \text{if}\;\;\;\;L<K_L\\
0  &\;\;\;\; \text{if}\;\;\;\; L \geqslant K_L.
\end{array}%
\right.
\end{eqnarray}
Therefore, if the number of larvae exceeds the larval capacity $K_L$
then no wet egg can make a transition to the larval compartment. On
the other hand, it is known that the larval population growth is
restricted by intraspecific competition, which increases the larval death
rate. In our model we include this effect by adding to the larval
mortality rate $\mu_L$ the following term~(see Ref.~\cite{otero2006stochastic}):
\begin{eqnarray}\label{eq.cl}
C_L=1.5 \frac{L}{K_L}.
\end{eqnarray}

\subsubsection{Mortality rates}\label{Ap.MorRat}
In Table~\ref{tab.Death} we define the mortality rates and their
relation with temperature (see Ref.~\cite{barmak2014modelling} and references therein).
\begin{table}[H]
\centering
\caption{Mortality rate coefficients}
\label{tab.Death}
\begin{tabular}{|c|p{6.5cm}|c|}
\hline
Quantity & Definition & Value  \\
\hline
$\mu_E$& egg mortality rate (days$^{-1}$)  &0.011\\
$\mu_L$& larva mortality rate (days$^{-1}$)  &$0.01+0.9725 \exp(-(T-4.85)/2.7035)$\\
$\mu_P$& pupa mortality rate (days$^{-1}$)  &$0.01+0.9725 \exp(-(T-4.85)/2.7035)$\\
$\mu_M$& mosquito mortality rate (days$^{-1}$)   &0.091\\
\hline
\end{tabular}
\end{table}

\section*{Bibliography}
\bibliographystyle{unsrt} 
\bibliography{bib}

\end{document}